\newif\ifcol
\coltrue

\newif\ifbw
\bwfalse

\newif\ifdraft
\draftfalse
\newcommand{\pagesize}{a4paper}
\ifdraft
  \colfalse
\fi

\newlength{\plotwidth}
\ifdraft
\fi

\ifdraft
  \documentclass[11pt,draftcls,\pagesize,onecolumn,twoside]{IEEEtran}
\else
  \documentclass[10pt,journal,\pagesize,twoside]{IEEEtran}
\fi

\usepackage{url}

\usepackage{ifpdf}

\ifpdf
  \usepackage[pdftex]{hyperref}
\else
  \usepackage[ps2pdf]{hyperref}
\fi

\usepackage{booktabs}

\usepackage{cite}      % Written by Donald Arseneau
\ifpdf
  \usepackage[pdftex]{graphicx}
\else
  \usepackage[dvips]{graphicx}
\fi

\usepackage{subfigure} % Written by Steven Douglas Cochran

\usepackage{amssymb}    % Required for some math symbols uses, e.g. R for real numbers.

\usepackage{amsmath}    % From the American Mathematical Society
\usepackage{algpseudocode}

\newcommand{\eqn}[1]{(#1)}

\newcommand{\fig}[1]{Fig.~#1}

\newcommand{\sectn}[1]{Sec.~#1}

\newcommand{\etal}{\mbox{\it et al.}}
\newcommand{\eg}{\mbox{\it e.g.}}
\newcommand{\ie}{\mbox{\it i.e.}}

\newcommand{\fft}{{FFT}}

\newcommand{\cmb}{{CMB}}
\newcommand{\cmbtext}{{cosmic microwave background}}

\newcommand{\healpix}{{\tt HEALPix}}

\newcommand{\spcend}{\ensuremath{\:}}
\newcommand{\img}{\ensuremath{{\rm i}}}
\newcommand{\cconj}{\ensuremath{\ast}}

\newcommand{\integers}{\ensuremath{\mathbb{Z}}}
\newcommand{\naturals}{\ensuremath{\mathbb{N}}}

\newcommand{\ltwo}{\ensuremath{\mathrm{L}^2}}
\newcommand{\sphere}{\ensuremath{{\mathrm{S}^2}}}
\newcommand{\sothree}{\ensuremath{{\mathrm{SO}(3)}}}
\newcommand{\torus}{\ensuremath{{\mathrm{T}^2}}}

\newcommand{\dx}{\ensuremath{\mathrm{\,d}}}
\newcommand{\dmu}[1]{\ensuremath{\dx \Omega(#1)}}

\newcommand{\innerp}[2]{\ensuremath{\langle {#1},\: {#2} \rangle}}

\newcommand{\saa}{\ensuremath{\theta}}
\newcommand{\sab}{\ensuremath{\varphi}}
\newcommand{\sas}{\ensuremath{\saa, \sab}}

\newcommand{\euls}{\ensuremath{\eula, \eulb, \eulc}}
\newcommand{\eula}{\ensuremath{\alpha}}
\newcommand{\eulb}{\ensuremath{\beta}}
\newcommand{\eulc}{\ensuremath{\gamma}}
\newcommand{\el}{\ensuremath{\ell}}
\newcommand{\m}{\ensuremath{m}}
\newcommand{\n}{\ensuremath{n}}
\newcommand{\spin}{\ensuremath{s}}

\newcommand{\elmax}{\ensuremath{{L}}}
\newcommand{\p}{\ensuremath{^\prime}}

\newcommand{\kron}[2]{\ensuremath{\delta_{{#1}{#2}}}}

\renewcommand{\exp}[1]{\ensuremath{{\rm e}^{#1}}}
\newcommand{\shfarg}[3]{\ensuremath{Y_{#1#2}({#3})}}
\newcommand{\shfargc}[3]{\ensuremath{Y_{#1#2}^\cconj({#3})}}
\newcommand{\shfargsp}[3]{\ensuremath{Y_{{#1},{#2}}({#3})}}
\newcommand{\sshfarg}[4]{\ensuremath{{{}_{#4} Y_{#1#2}({#3})}}}
\newcommand{\sshfargc}[4]{\ensuremath{{{}_{#4} Y_{#1#2}^\cconj({#3})}}}
\newcommand{\sshfargsp}[4]{\ensuremath{{{}_{#4} Y_{{#1},{#2}}({#3})}}}
\newcommand{\shf}[2]{\ensuremath{Y_{#1#2}}}

\newcommand{\shc}[3]{\ensuremath{{#1}_{{#2}{#3}}}}
\newcommand{\shcc}[3]{\ensuremath{{#1}_{{#2}{#3}}^\cconj}}
\newcommand{\shcsp}[3]{\ensuremath{{#1}_{{#2},{#3}}}}

\newcommand{\sshf}[3]{\ensuremath{{{}_{#3} Y_{#1#2}}}}

\newcommand{\aleg}[3]{\ensuremath{P_{#1}^{#2}({#3})}}

\newcommand{\jacobi}[4]{\ensuremath{P_{#1}^{(#2,#3)}({#4})}}

\newcommand{\dmatbig}{\ensuremath{D}}
\newcommand{\Dlmn}{\ensuremath{ \dmatbig_{\m\n}^{\el} }}

\newcommand{\dmatsmall}{\ensuremath{d}}
\newcommand{\dlmn}{\ensuremath{ \dmatsmall_{\m\n}^{\el} }}

\newcommand{\dlmnb}{\ensuremath{ \dmatsmall_{\m\n}^{\el}(\eulb) }}

\newcommand{\dlmnhalfpi}[3]{\ensuremath{ \Delta_{{#2}{#3}}^{#1} }}

\newcommand{\dlmnhalfpim}{\ensuremath{ \Delta_{{\m\p}{\m}}^{\el} }}
\newcommand{\dlmnhalfpisn}{\ensuremath{ \Delta_{{\m\p}{,-\spin}}^{\el} }}

\newcommand{\spinup}{\ensuremath{\eth}}
\newcommand{\spindown}{\ensuremath{\bar{\eth}}}

\newcommand{\f}{\ensuremath{f}}
\newcommand{\fs}{\ensuremath{{}_\spin f}}
\newcommand{\fsm}{\ensuremath{{{}_{-\spin} f}}}
\newcommand{\flm}{\ensuremath{\shc{\f}{\el}{\m}}}
\newcommand{\fslm}{\ensuremath{\shc{\fs}{\el}{\m}}}

\newcommand{\elmfact}{\ensuremath{\frac{(\el-\m)!}{(\el+\m)!}}}

\newcommand{\sumlm}{\ensuremath{\sum_{\el=0}^{\infty} \sum_{\m=-\el}^\el}}

\newcommand{\summp}{\ensuremath{\sum_{\m\p=-\el}^\el}}

\newcommand{\nl}{\ensuremath{\sqrt{\frac{2\el+1}{4\pi}}}}

\newcommand{\G}[3]{\ensuremath{{{}_{#1} G_{{#2} {#3}}}}}

\newcommand{\rG}[3]{\ensuremath{{{}_{#1} \tilde{G}_{{#2} {#3}}}}}
\newcommand{\Gsm}{\ensuremath{\G{\spin}{\m}{}}}
\newcommand{\Gsmm}{\ensuremath{\G{\spin}{\m}{\m\p}}}
\newcommand{\Gsmt}{\ensuremath{\G{\spin}{\m}{}(\saa)}}
\newcommand{\rGsmt}{\ensuremath{\rG{\spin}{\m}{}(\saa)}}

\newcommand{\Gsmti}{\ensuremath{\G{\spin}{\m}{}({\saaiang})}}

\newcommand{\rGsmti}{\ensuremath{\rG{\spin}{\m}{}({\saaiang})}}
\newcommand{\F}[3]{\ensuremath{{{}_{#1} F_{{#2} {#3}}}}}

\newcommand{\Fsm}{\ensuremath{\F{\spin}{\m}{}}}
\newcommand{\Fsmm}{\ensuremath{\F{\spin}{\m}{\m\p}}}
\newcommand{\Fzmm}{\ensuremath{\F{0}{\m}{\m\p}}}
\newcommand{\Fsmmn}{\ensuremath{\F{\spin}{\m}{,-\m\p}}}

\newcommand{\Fzmnmn}{\ensuremath{\F{0}{-\m}{,-\m\p}}}
\newcommand{\Fsmmp}{\ensuremath{\F{\spin}{\m}{\m{\p}{\p}}}}
\newcommand{\Fsmt}{\ensuremath{\F{\spin}{\m}{}(\saa)}}

\newcommand{\Fsmtn}{\ensuremath{\F{\spin}{\m}{}(-\saa)}}

\newcommand{\intsaa}{\ensuremath{\int_0^{\pi} \dx \saa \sin \saa}}
\newcommand{\intsab}{\ensuremath{\int_0^{2\pi} \dx \sab}}
\newcommand{\saai}{\ensuremath{t}}
\newcommand{\sabi}{\ensuremath{p}}
\newcommand{\saaiang}{\ensuremath{\saa_\saai}}
\newcommand{\sabiang}{\ensuremath{\sab_\sabi}}

\newcommand{\saisang}{\ensuremath{\saaiang,\sabiang}}
\newcommand{\sumsaai}{\ensuremath{\sum_{\saai=-(\elmax-1)}^{\elmax-1}}}
\newcommand{\sumsabi}{\ensuremath{\sum_{\sabi=-(\elmax-1)}^{\elmax-1}}}
\newcommand{\sumltrunc}{\ensuremath{\sum_{\el=0}^{\elmax-1}}}
\newcommand{\summtrunc}{\ensuremath{\sum_{\m=-(\elmax-1)}^{\elmax-1}}}
\newcommand{\summptrunc}{\ensuremath{\sum_{\m\p=-(\elmax-1)}^{\elmax-1}}}
\newcommand{\qweight}{\ensuremath{q}}
\newcommand{\weighttrans}{\ensuremath{v}}

\newcommand{\N}{\ensuremath{{N}}}
\newcommand{\Ngl}{\ensuremath{{N_{\rm GL}}}}
\newcommand{\Ndh}{\ensuremath{{N_{\rm DH}}}}
\newcommand{\Nhw}{\ensuremath{{N_{\rm HW}}}}
\newcommand{\Nmw}{\ensuremath{{N_{\rm MW}}}}
\newcommand{\Nm}{\ensuremath{{N_{\rm M}}}}
\newcommand{\Ns}{\ensuremath{{N_{\rm S}}}}

\newcommand{\weight}{\ensuremath{w}}
\newcommand{\order}{\ensuremath{\mathcal{O}}}

\begin{document}
\title{A novel sampling theorem on the sphere}

% 

\author{Jason~D.~McEwen~and~Yves~Wiaux
%        Michael~Shell,~\IEEEmembership{Member,~IEEE,}
%        John~Doe,~\IEEEmembership{Fellow,~OSA,}
%        and~Jane~Doe,~\IEEEmembership{Life~Fellow,~IEEE}% <-this % stops a space
\thanks{Manuscript received March 23, 2011.}%
\thanks{The work of
  J.~D.~McEwen is supported by the Swiss National Science Foundation
  (SNSF) under grant 200021-130359.  The work of Y.~Wiaux is supported
  in part by the Center for Biomedical Imaging (CIBM) of the Geneva
  and Lausanne Universities, EPFL, and the Leenaards and Louis-Jeantet
  foundations, and in part by the SNSF under grant PP00P2-123438.}%
\thanks{The authors are with the the Institute of Electrical
  Engineering, Ecole Polytechnique F{\'e}d{\'e}rale de Lausanne
  (EPFL), CH-1015 Lausanne, Switzerland.  Y.~Wiaux is also with the
  Institute of Bioengineering, EPFL, \mbox{CH-1015} Lausanne,
  Switzerland, and the Department of Radiology and Medical
  Informatics, University of Geneva (UniGE), CH-1211 Geneva,
  Switzerland.}%
\thanks{E-mail: jason.mcewen@ucl.ac.uk (J.~D.~McEwen)}}
% note the % following the last \IEEEmembership and also the first \thanks - 
% instance, "A\textbf{} \textbf{}B" will typeset as "A B" not "AB". If you want
% "AB" then you have to do: "A\textbf{}\textbf{}B"
\markboth{IEEE Transactions on Signal Processing,~Vol.~59, No.~12,~December~2011}%
{McEwen \& Wiaux: A novel sampling theorem on the sphere}
% *** Note that you probably will NOT want to include the author's name in ***
% *** the headers of peer review papers.                                   ***

% If you want to put a publisher's ID mark on the page
% (can leave text blank if you just want to see how the
% text height on the first page will be reduced by IEEE)
%\pubid{0000--0000/00\$00.00~\copyright~2002 IEEE}

%\specialpapernotice{(Invited Paper)}

\maketitle

%------------------------------------------------------------------------------

% Abstract.
%------------------------------------------------------------------------------
% Abstract
%------------------------------------------------------------------------------

\begin{abstract}  
We develop a novel sampling theorem on the sphere and corresponding fast algorithms by associating the sphere with the torus through a periodic extension.  The fundamental property of any sampling theorem is the number of samples required to represent a band-limited signal.  To represent exactly a signal on the sphere band-limited at $\elmax$, all sampling theorems on the sphere require $\order(\elmax^2)$ samples.  However, our sampling theorem requires less than half the number of samples of other equiangular sampling theorems on the sphere and an asymptotically identical, but smaller, number of samples than the Gauss-Legendre sampling theorem.  The complexity of our algorithms scale as $\order(\elmax^3)$, however, the continual use of fast Fourier transforms reduces the constant prefactor associated with the asymptotic scaling considerably, resulting in algorithms that are fast.  Furthermore, we do not require any precomputation and our algorithms apply to both scalar and spin functions on the sphere without any change in computational complexity or computation time.  We make our implementation of these algorithms available publicly and perform numerical experiments demonstrating their speed and accuracy up to very high band-limits.  Finally, we highlight the advantages of our sampling theorem in the context of potential applications, notably in the field of compressive sampling.
\end{abstract}

% ctrl+c _
%%% Local Variables: 
%%% mode: latex
%%% TeX-master: "fssht"
%%% End: 3

% Keywords.

% Will be ignored for other modes.
\IEEEpeerreviewmaketitle

% Main body.
%==============================================================================
\section{Introduction}
%==============================================================================

\PARstart{I}{n} many fields of science and engineering data are measured on a spherical manifold.  Applications where data are defined inherently on the sphere are found in computer graphics (\eg\ \cite{ramamoorthi:2004}), planetary science (\eg\ \cite{turcotte:1981,wieczorek:2006,wieczorek:1998,audet:2010}), geophysics (\eg\ \cite{simons:2006,swenson:2002,whaler:1994}), quantum chemistry (\eg\ \cite{choi:1999,ritchie:1999}) and astrophysics (\eg\ \cite{bennett:1996,jaroski:2010}%
%bennett:2003a,hinshaw:2006,hinshaw:2008}
), to quote only a few.  In many of these applications a harmonic analysis of the data is insightful.  For example, spherical harmonic analyses have been remarkably successful in cosmology over the past decade, leading to the emergence of a standard cosmological model.  Observations of the anisotropies of the \cmbtext\ (\cmb), which are made on the celestial sphere, contain a wealth of information about the early Universe.  Cosmologists extract this information from the angular power spectrum of observations of the CMB, computed through a harmonic transform on the sphere (\eg\ \cite{komatsu:2010}).  Recent and upcoming full-sky observations of the \cmb\ are of considerable size, containing approximately three \cite{jaroski:2010} and fifty \cite{planck:bluebook} million samples respectively.  Furthermore, observations are made of both the temperature and polarisation of the \cmb, which give rise to scalar and spin $\pm 2$ functions on the sphere respectively.  Consequently, the ability to perform fast scalar and spin spherical harmonic transforms on large data sets is of considerable importance in cosmology and beyond.

Algorithms to perform spherical harmonic transforms have received considerable attention already.  Some correspond to sampling theorems on the sphere, where the forward and inverse transform are theoretically exact for a band-limited signal on the sphere.  Others adopt approximate quadrature rules on the sphere, resulting in approximate harmonic transforms that do not correspond to sampling theorems on the sphere.  However, these approximate algorithms typically arise from particular pixelisations of the sphere which meet desirable practical criteria, such as pixels of equal area, and are no less important.  We focus here on approaches that lead to sampling theorems on the sphere with theoretically exact transforms for signals on the sphere band-limited at \elmax\ (where the harmonic band-limit \elmax\ is defined formally in \sectn{\ref{sec:background:spin}}).  Note that the current \cite{jaroski:2010} and forthcoming \cite{planck:bluebook} \cmb\ observations discussed previously support band-limits of $\elmax=1024$ and $\elmax=4096$ respectively.  All sampling theorems require $\N$ samples on the sphere of order $\order(\elmax^2)$, however the exact number of samples required varies for each sampling theorem.  For many applications reducing the number of samples required to represent a band-limited signal on the sphere is of fundamental importance.

Sampling theorems on the sphere and their associated numerical algorithms are evaluated by four criteria:  (i) the number of samples required to represent a band-limited signal exactly; (ii) their computational complexity; (iii) their speed; and (iv) issues surrounding any precomputation.  From an information theoretic viewpoint, the fundamental property of any sampling theorem is the number of samples required to represent a band-limited signal exactly.  %For example, in Euclidean space the Shannon sampling theorem \cite{shannon:1949} gives rise to the optimal Nyquist sampling.
In this article we review algorithms to compute spherical harmonic transforms accurately and efficiently.  We also present a novel sampling theorem on the sphere and corresponding fast algorithms.  Our approach compares favourably to the state-of-the-art as evaluated by the four criteria listed previously.  Furthermore, our algorithms apply to both scalar and spin functions on the sphere without any change in asymptotic complexity or computation time.

The remainder of this article is structured as follows.  In \sectn{\ref{sec:review}} we review comprehensively the literature regarding the computation of spherical harmonic transforms, placing our novel sampling theorem and fast algorithms in the context of preceding work.  Harmonic analysis on the sphere is reviewed concisely in \sectn{\ref{sec:background}}, to present the mathematical preliminaries required subsequently.  Our sampling theorem on the sphere and the corresponding fast algorithms to compute spherical harmonic transforms are derived in \sectn{\ref{sec:fsht}}.  In \sectn{\ref{sec:evaluation}} we evaluate our algorithms numerically and discuss the advantages of our sampling theorem in the context of potential applications, notably in the field of compressive sampling. Concluding remarks are made in \sectn{\ref{sec:conclusions}}.

%==============================================================================
\section{Review}
\label{sec:review}
%==============================================================================

The development of sampling theorems on the sphere and fast algorithms to compute spherical harmonic transforms has been driven largely by researchers in the fields of computational harmonic analysis, geophysics and astrophysics.  Due to the diverse nature of these fields, the literature on this topic appears to be somewhat disjoint.  We attempt to unify these works here and to present a comprehensive review of the historical development of the field.  

For isolatitude sampling schemes, where the samples are gathered in isolatitude annuli, a separation of variables may be performed to rewrite the scalar spherical harmonic transform as a Fourier transform in longitude and an associated Legendre transform in colatitude.  The computation is then dominated by the associated Legendre transform, reducing the complexity from $\order(\elmax^4)$ to $\order(\elmax^3)$.  Isolatitude samplings are thus very common and hence we restrict our attention to such schemes.  Spin lowering and raising operators can be used to relate the harmonic transform of spin functions to the scalar case, hence scalar transforms have received the majority of attention.  Approaches to improve the computational performance of scalar spherical harmonic transforms attempt either to reduce the asymptotic complexity further through fast associated Legendre transforms or to reduce the constant prefactor associated with the asymptotic scaling of the algorithm.

First attempts to compute a scalar spherical harmonic transform through a fast Legendre transform were performed by Orszag \cite{orszag:1986} and were based on a Wentzel-Kramers-Brillouin (WKB) approximation.  Alternative approaches using the fast multipole method (FMM) \cite{beatson:1997} have been considered by Alpert \& Rokhin \cite{alpert:1991} and by Suda \& Takani \cite{suda:2002}.  The complexity of these algorithms scale linearly with the desired accuracy.  An alternative approximate algorithm using WKB frequency estimates has been developed by Mohlenkamp \cite{mohlenkamp:1997,mohlenkamp:1999} for functions with band-limits that are a power of two, however this approximation can be controlled independently of complexity, which scales as $\order(\elmax^2 \log {}_2^2 \elmax)$.  In any case, these types of approach are necessarily approximate and do not yield exact sampling theorems on the sphere.

Other approximate approaches based solely on the separation of variables have been developed more recently for pixelisations of the sphere that satisfy certain practical requirements, such as \healpix\footnote{\url{http://healpix.jpl.nasa.gov/}} \cite{gorski:2005} and {\tt IGLOO}\footnote{\url{http://www.mrao.cam.ac.uk/projects/cpac/igloo/}} \cite{crittenden:1998}, resulting in algorithms of complexity $\order(\elmax^3)$.  For these pixelisations only approximate quadrature rules exist, hence the spherical harmonic transform algorithms of \healpix\ and {\tt IGLOO} are not theoretically exact.  Nevertheless, these pixelisation schemes satisfy a number of desirable practical criteria, such as pixels of equal area, and their associated harmonic transform algorithms are of sufficient accuracy for many practical purposes.  These schemes have found considerable application in the analysis of \cmb\ data.

Exact transforms with associated sampling theorems have been constructed for particular pixelisation schemes.  It is well-known that Gauss-Legendre quadrature may be used to construct exact spherical harmonic transforms.  To our knowledge, this result was first highlighted in published work by Shukowsky \cite{shukowsky:1986}, which in turn refers to unpublished (and inaccessible) work by Payne from 1971 \cite{payne:1971}.  An exact sampling theorem can be constructed from \mbox{$\Ngl = \elmax (2 \elmax-1) \sim 2 \elmax^2$} samples on the sphere, where the sample locations in colatitude are chosen as the roots of the Legendre polynomials of order \elmax, as dictated by Gauss-Legendre quadrature.  Through a separation of variables, the resulting algorithm is $\order(\elmax^3)$.  The {\tt GLESP}\footnote{\url{http://www.glesp.nbi.dk/}} \cite{doroshkevich:2005} pixelisation scheme has been constructed using Gauss-Legendre quadrature, however this scheme uses twice as many samples in colatitude as required, \ie\ approximately $2 \Ngl \sim 4 \elmax^2$ samples are used.  {\tt GLESP} has also found considerable application in the analysis of \cmb\ data.

The first theoretically exact sampling theorem on an equiangular pixelisation was developed by Shukowsky \cite{shukowsky:1986}, requiring \mbox{$\Ns = (2 \elmax-1)^2 \sim 4 \elmax^2$} samples on the sphere; however, the exactness of this approach was not studied numerically. Although this algorithm remains $\order(\elmax^4)$, a separation of variables may be used to reduce the computational complexity to $\order(\elmax^3)$. An alternative sampling theorem on the sphere for an equiangular pixelisation was developed by Driscoll and Healy \cite{driscoll:1994}.
Moreover, a divide-and-conquer approach to computing a fast associated Legendre transform in the cosine basis was derived \cite{driscoll:1994}.  The resulting algorithm is exact in exact precision arithmetic, and has computational complexity $\order(\elmax^2 \log {}_2^2\elmax)$, but is known to suffer from stability problems \cite{healy:2003,kostelec:2008}.  %Potts \cite{potts:1998} adapted the Driscoll \& Healy approach to use a fast cosine transform (as suggested in \cite{driscoll:1994}) to reduce computation time (but not complexity) and introduced some stabilisation methods.  
Healy \etal\ \cite{healy:2003} readdressed the work of Driscoll \& Healy \cite{driscoll:1994}, reformulating the sampling theorem on the sphere and developing some variants of the original algorithm,\footnote{
Healy \etal\ \cite{healy:2003} derive a number of variants of the original Driscoll \& Healy algorithm \cite{driscoll:1994}, including the so-called \emph{semi-naive}, \emph{simple-split} and \emph{hybrid} algorithms.  The semi-naive algorithm avoids dividing (and conquering) the problem, resulting in $\order(\elmax^3)$ complexity.  The simple-split algorithm is a  simpler and more stable divide-and-conquer approach than the original algorithm but with an increased complexity of $\order(\elmax^{5/2} \log {}_2^{1/2} \elmax)$ and is less stable than the semi-naive approach.  The splitting required by the simple-split algorithm is costly (in terms of execution time rather than asymptotic complexity), thus for a band-limit of $\elmax=1024$ the semi-naive algorithm is greater than two times faster than the simple-split algorithm \cite{healy:2003}.  The hybrid algorithm attempts to mitigate the slow execution of the simple-spit algorithm and the higher complexity of the semi-naive algorithm by splitting the problem between them.  The hybrid algorithm appears to achieve a good compromise between stability and efficiency.  However, the overall complexity of this algorithm is not clear since it depends on the split between the semi-naive and simple-split algorithms and on user specified parameters.} which are available for download.\footnote{\url{http://www.cs.dartmouth.edu/~geelong/sphere/}}  However, the only variant that is universally stable is the so-called \emph{semi-naive} algorithm, which remains $\order(\elmax^3)$.  Algorithms to compute \mbox{spin $\pm2$} transforms are derived by Wiaux \etal\ \cite{wiaux:2005b} and Kostelec \etal\ \cite{kostelec:2000} using spin raising and lowering operators to relate spin transforms to the scalar case, before applying the scalar algorithms developed by Healy \etal\ \cite{healy:2003}.  In general, this type of approach may be used to compute spin transforms for arbitrary spin \cite{wiaux:2005c}, however the complexity of the resulting algorithm then also scales linearly with spin number.
All of these algorithms \cite{driscoll:1994,potts:1998,healy:2003,wiaux:2005b,wiaux:2005c} require \mbox{$\Ndh = 2 \elmax (2 \elmax-1) \sim 4 \elmax^2$} samples on the sphere and, moreover, the divide-and-conquer based approaches are all restricted to harmonic band-limits that are a power of two.  Furthermore, all of the methods with complexity below $\order(\elmax^3)$ require a precomputation that requires $\order(\elmax^3)$ computations and storage.  At band-limit \mbox{$\elmax=1024$}, for example, the precomputation requires $1.2$GB of storage \cite{wiaux:2005b}, scaling to approximately $77$GB for the band-limit $\elmax=4096$ of forthcoming \cmb\ observations.   Precomputation quickly becomes infeasible for high band-limits, thus the $\order(\elmax^3)$ semi-naive algorithm is the most universally applicable fast algorithm implementing the Driscoll \& Healy sampling theorem.

Other approaches to reduce the cost of computing spherical harmonic transforms focus on reducing the constant prefactor associated with the asymptotic complexity of algorithms.  These approaches have typically exploited fast Fourier transforms (\fft s) on equiangular pixelisations to reduce computation time through an association between the sphere and the torus, while their complexity remains $\order(\elmax^3)$.  To our knowledge, the first algorithm based on this technique was developed by Dilts \cite{dilts:1985}, where the North and South poles of the sphere were identified to map the sphere to the torus.  However, this algorithm is approximate and does not result in a sampling theorem on the sphere.  One of the authors of this article developed a sampling theorem on the sphere \cite{mcewen:fsht} by making periodic extensions of the sphere in colatitude in order to make an association with the torus.  This sampling theorem requires \mbox{$\Nm = (\elmax - 1) (2 \elmax - 1) + 1 \sim 2 \elmax^2$} samples on the sphere and, moreover, applies to any spin number
 without the application of spin lowering and raising operators.  However, the forward algorithm associated with this sampling theorem proved unstable (the inverse algorithm did not suffer from stability issues).  Recently, Huffenberger \& Wandelt \cite{huffenberger:2010} adopted the inverse transform of this approach and resolved the instability in the forward algorithm, although in doing so increased the number of points required to sample a band-limited function on the sphere to \mbox{$\Nhw = 2\elmax (2 \elmax - 1) \sim 4 \elmax^2$}.  
In this article we readdress sampling theorems derived by associating the sphere with the torus through periodic extensions and develop a sampling theorem requiring \mbox{$\Nmw = (\elmax - 1) (2 \elmax - 1) + 1 \sim 2 \elmax^2$} samples on an equiangular pixelisation, with corresponding fast algorithms that do \emph{not} suffer from any stability issues.

%==============================================================================
\section{Harmonic Analysis on the Sphere}
\label{sec:background}
%==============================================================================

In this section we review harmonic analysis on the two-sphere \sphere.  We first review the scalar spherical harmonic transform, before generalising to the spin case.  Associations are then made between the spin spherical harmonics and the Wigner functions, where the latter provide an orthogonal basis for the decomposition of square integrable functions on the rotation group \sothree.

%==============================================================================
\subsection{Scalar spherical harmonics}
\label{sec:background:scalar}

We consider the space of square integrable functions on the sphere $\ltwo(\sphere)$, with the inner product of $f,g\in\ltwo(\sphere)$ defined by
\begin{equation*}
\innerp{f}{g} = \int_\sphere \dmu{\sas} \: f(\sas) \: g^\cconj(\sas) 
\spcend ,
\end{equation*}
where $\dmu{\sas} = \sin \saa \dx \saa \dx \sab$ is the usual invariant measure on the sphere and $(\sas)$ define spherical coordinates with colatitude $\saa \in [0,\pi]$ and longitude $\sab \in [0,2\pi)$.   Complex conjugation is denoted by the superscript ${}^\cconj$.  

The scalar spherical harmonic functions form the canonical orthogonal basis for the space of $\ltwo(\sphere)$ scalar functions on the sphere and are defined by
\begin{equation*}
%\label{eqn:shf}
\shfarg{\el}{\m}{\sas} = 
\sqrt{\frac{2\el+1}{4\pi} \elmfact} \:
\aleg{\el}{\m}{\cos\saa} \:
\exp{\img \m \sab} 
\spcend ,
\end{equation*}
for natural $\el\in\naturals$ and integer $\m\in\integers$, $|\m|\leq\el$, where $\aleg{\el}{\m}{x}$ are the associated Legendre functions. 
We adopt the Condon-Shortley phase convention, with the $(-1)^\m$ phase factor included in the definition of the associated Legendre functions, to ensure that the conjugate symmetry relation $\shfargc{\el}{\m}{\sas} = (-1)^\m \: \shfargsp{\el}{-\m}{\sas}$ holds.
The orthogonality and completeness relations for the spherical harmonics read
%\begin{equation*}
%\label{eqn:shortho}
$
\innerp{\shf{\el}{\m}}{\shf{\el\p}{\m\p}}
= 
\kron{\el}{\el\p}
\kron{\m}{\m\p}
$
%\end{equation*}
and
\begin{equation*}
\sumlm
\shfarg{\el}{\m}{\saa,\sab} \:
\shfargc{\el}{\m}{\saa\p,\sab\p} 
=
\delta(\cos\saa - \cos\saa\p) \:
\delta(\sab - \sab\p)
\end{equation*}
respectively, where $\kron{i}{j}$ is the Kronecker delta symbol and $\delta(x)$ is the Dirac delta function.

Due to the orthogonality and completeness of the scalar spherical harmonics, any square integrable scalar function on the sphere $\f \in \ltwo(\sphere)$ may be represented by its spherical harmonic expansion
\begin{equation*}
%\label{eqn:sht_inv}
\f(\sas) = 
\sum_{\el=0}^\infty
\sum_{\m=-\el}^\el
\flm \:
\shfarg{\el}{\m}{\sas}
\spcend ,
\end{equation*}
where the spherical harmonic coefficients are given by the usual projection onto each basis function:
$\shc{\f}{\el}{\m} =
\innerp{f}{\shf{\el}{\m}}$.
The conjugate symmetry relation of the spherical harmonic coefficients of a real function is given by \mbox{$\shcc{f}{\el}{\m} = (-1)^\m \: \shcsp{f}{\el}{-\m}$}, which follows directly from the conjugate symmetry of the scalar spherical harmonic functions. 

%==============================================================================
\subsection{Spin spherical harmonics}
\label{sec:background:spin}

Square integrable spin functions on the sphere $\fs\in \ltwo(\sphere)$, with integer spin $\spin\in\integers$, are defined by their behaviour under local rotations.  By definition, a spin function transforms as
\begin{equation}
\label{eqn:spin_rot}
\fs^\prime(\sas) = \exp{-\img \spin \chi} \: \fs(\sas)
\end{equation}
under a local rotation by $\chi$, where the prime denotes the rotated function.  It is important to note that the rotation considered here is \emph{not} a global rotation on the sphere, such as that represented by an element of the rotation group \sothree, but rather a rotation by $\chi$ in the tangent plane at $(\sas)$.  The sign convention that we adopt here for the argument of the complex exponential in \eqn{\ref{eqn:spin_rot}} differs to the original definition \cite{newman:1966} but is identical to the convention used recently in the context of the polarisation of the \cmb\ \cite{zaldarriaga:1997}.

The spin spherical harmonics $\sshfarg{\el}{\m}{\sas}{\spin}$ form an orthogonal basis for $\ltwo(\sphere)$ spin \spin\ functions on the sphere for $|\spin|\leq\el$.  Spin spherical harmonics were first developed by Newman \& Penrose \cite{newman:1966} and were soon realised by Goldberg \cite{goldberg:1967} to be closely related to the Wigner functions.  We therefore defer the explicit definition of the spin spherical harmonic functions until \sectn{\ref{sec:background:wigner}}.
The conjugate symmetry relation given for the spin spherical
harmonics is given by
\mbox{$\sshfargc{\el}{\m}{\sas}{\spin} = (-1)^{\spin+\m}
\sshfargsp{\el}{-\m}{\sas}{-\spin}$}.
The spin spherical harmonics satisfy identical orthogonality and completeness relations as the scalar spherical harmonics.
% orthogonality and completeness of the spin spherical
% harmonics reads
% %\begin{equation}
% $
% \innerp{\sshf{\el}{\m}{\spin}}
% {\sshf{\el\p}{\m\p}{\spin}}
%  = 
% \kron{\el}{\el\p}
% \kron{\m}{\m\p}
% $
% %\end{equation}
% and
% \begin{equation*}
% \sumlm
% \sshfarg{\el}{\m}{\saa,\sab}{\spin} \:
% \sshfargc{\el}{\m}{\saa\p,\sab\p}{\spin}
% =
% \delta(\cos\saa - \cos\saa\p) \:
% \delta(\sab - \sab\p)
% \end{equation*}
% respectively.

Due to the orthogonality and completeness of the spin spherical harmonics, any square integrable spin function on the sphere $\fs \in \ltwo(\sphere)$ may be represented by its spherical harmonic expansion
\begin{equation*}
\fs(\sas) = 
\sum_{\el=0}^\infty
\sum_{\m=-\el}^\el
\fslm \:
\sshfarg{\el}{\m}{\sas}{\spin}
\spcend ,
\end{equation*}
where the spin spherical harmonic coefficients are given by the usual projection onto each basis function: $\fslm = \innerp{\fs}{\sshf{\el}{\m}{\spin}}$.  Note that the spin spherical harmonics and transforms simply generalise the scalar equivalents to spin signals, reducing to the standard scalar case for $\spin=0$.
When deriving our novel sampling theorem we consider signals on the sphere band-limited at $\elmax$, that is signals such that $\shc{\fs}{\el}{\m}=0$, $\forall \el\geq\elmax$. 
The conjugate symmetry relation of the spin spherical harmonic coefficients is given by $\shcc{\fs}{\el}{\m} = (-1)^{\spin+\m} \: {}_{-\spin}\shcsp{\f}{\el}{-\m}$ for a function satisfying \mbox{$\fs^\cconj=\fsm$} (which for a spin $\spin=0$ function equates to the usual reality condition) and follows directly from the conjugate symmetry of the spin spherical harmonics.

Spin raising and lowering operators, $\spinup$ and $\spindown$
respectively, exist so that spin $\spin\pm1$ functions may be obtained
from spin \spin\ functions \cite{newman:1966,goldberg:1967}.  Spin raising and lowering operators are often used repeatedly to relate spin $\spin$ functions to scalar functions on the sphere.

%==============================================================================
\subsection{Wigner functions}
\label{sec:background:wigner}

The Wigner functions $\Dlmn(\euls)$, for natural $\el \in \naturals$ and integer $\m,\n \in \integers$, form an orthogonal basis for the space $\ltwo(\sothree)$ of square integrable functions on the rotation group,
and are parameterised by the Euler angles $(\euls)$, where $\eula \in [0,2\pi)$, $\eulb \in [0,\pi]$ and $\eulc \in [0,2\pi)$.\footnote{We adopt the $zyz$ Euler convention corresponding to the rotation of a physical body in a \emph{fixed} co-ordinate system about the $z$, $y$ and $z$ axes by \eulc, \eulb\ and \eula\ respectively.}  
The Wigner functions may be decomposed as \cite{varshalovich:1989}%\cite{brink:1999,ritchie:1999}
\begin{equation}
\label{eqn:d_decomp}
\dmatbig_{\m\n}^{\el}(\euls)
= {\rm e}^{-\img \m\eula} \:
\dmatsmall_{\m\n}^\el(\eulb) \:
{\rm e}^{-\img \n\eulc}
\spcend ,
\end{equation}
where the real polar \dmatsmall-functions are defined by \cite{varshalovich:1989}
\begin{align}
\label{eqn:wignerd_b}
  \dlmnb =&
  \sqrt{\frac{(\el+\n)! (\el-\n)!}{(\el+\m)! (\el-\m)!}}
  \left( \sin\frac{\eulb}{2} \right)^{\n-\m} \nonumber \\ & \times
  \left( \cos\frac{\eulb}{2} \right)^{\n+\m} 
  \jacobi{\el-\n}{\n-\m}{\n+\m}{\cos\eulb}
  \spcend ,
\end{align}
where $\jacobi{\el}{a}{b}{\cdot}$ are the Jacobi polynomials.  Note that recursion formulae are available to compute rapidly the Wigner \mbox{\dmatsmall-functions} (\eg\ \cite{risbo:1996,trapani:2006}).  The \dmatsmall-functions satisfy a number of symmetry relations; in this work we make use of the symmetry relations \cite{varshalovich:1989}
\begin{equation}
  \label{eqn:wignerd_sym1} 
  \dlmnb = (-1)^{\m-\n} \:  \dmatsmall_{-\m,-\n}^\el(\eulb) \spcend  
  \spcend ,
\end{equation}
\begin{equation}
  \label{eqn:wignerd_sym2} 
  \dlmn(\pi-\eulb) = (-1)^{\el-\n} \: \dmatsmall_{-\m,\n}^\el(\eulb)
\end{equation}
and
\begin{equation}
  \label{eqn:wignerd_sym3} 
  \dlmn(-\eulb) = (-1)^{\m-\n} \: \dlmn(\eulb)
  \spcend .
\end{equation}

We are not concerned with decompositions of functions on the rotation group in this article but rather representations of the spherical harmonics by Wigner functions.  The spin spherical harmonics may be defined by the Wigner functions through \cite{goldberg:1967}
\begin{equation}
\label{eqn:ssh_wigner}
\sshfarg{\el}{\m}{\sas}{\spin} = (-1)^\spin 
\sqrt{\frac{2\el+1}{4\pi} } \:
\dmatbig_{\m,-\spin}^{\el\:\cconj}(\sab,\saa  ,0)
\spcend .
\end{equation}
Defining the spherical harmonics in this manner allows us to apply standard Wigner function decompositions to the spherical harmonic functions.  We subsequently make considerable use of the Fourier series decomposition of the \dmatsmall-functions given by 
\cite{nikiforov:1991}:
\begin{equation}
  \label{eqn:wigner_sum_reln}
  \dlmnb = \img^{\n-\m} \sum_{\m\p=-\el}^\el
  \dlmnhalfpi{\el}{\m\p}{\m} \:
  \dlmnhalfpi{\el}{\m\p}{\n} \:
  \exp{\img \m\p \eulb}
  \spcend .
\end{equation}
where \mbox{$\dlmnhalfpi{\el}{\m}{\n} \equiv \dlmn (\pi/2)$}.  This expression follows from a factoring of rotations as highlighted by Risbo \cite{risbo:1996}.
The Fourier series representation of \dlmnb\ given by \eqn{\ref{eqn:wigner_sum_reln}} allows one to write the spherical harmonic expansion of \fs\ in terms of a Fourier series expansion of \fs\ extended appropriately to the two-torus \torus\ (as discussed in more detail in \sectn{\ref{sec:fsht}}).  Consequently, \eqn{\ref{eqn:wigner_sum_reln}} is fundamental to the derivation of our sampling theorem on the sphere and fast algorithms.

%==============================================================================
\section{Fast Spherical Harmonic Transform}
\label{sec:fsht}
%==============================================================================

We derive fast algorithms for performing forward and inverse spin spherical harmonic transforms and discuss the corresponding sampling theorem on the sphere.  Our approach involves an extension of the sphere to the torus so that \fft s may be exploited to reduce the cost of computation.  It is related closely to the algorithms derived by one of the authors in a previous work \cite{mcewen:fsht} and to the algorithms derived by Huffenberger \& Wandelt \cite{huffenberger:2010}, however it does not suffer from the instabilities of the former approach and requires only half as many samples on the sphere as the latter approach.  We first present the general harmonic formulation of our algorithms, followed by a discussion of periodisation and discretisation, before deriving algorithms to perform exact forward and inverse transforms.  Our sampling theorem also leads to a new quadrature rule on the sphere, which we then present, followed by a discussion of symmetries that may be exploited to improve the efficiency of computation for real signals.

%==============================================================================
\subsection{Harmonic formulation}
\label{sec:fsht:formulation}

We consider the harmonic transform of spin functions on the sphere $\fs\in\ltwo(\sphere)$, band-limited at \elmax; consequently, all summations over or up to \el\ are truncated to $\elmax-1$.  Furthermore, harmonic coefficients are not defined for $|\m|>\el$, hence we define them to be zero to enforce the contraint  $|\m| \leq \el$ when summations are interchanged.

By noting the definition of the spin spherical harmonics in terms of Wigner functions \eqn{\ref{eqn:ssh_wigner}}, the Wigner decomposition \eqn{\ref{eqn:d_decomp}} and the Fourier expansion of the Wigner $\dmatsmall$-functions \eqn{\ref{eqn:wigner_sum_reln}}, the forward transform of \fs\ may be written
\begin{equation}
  \label{eqn:flm_wig}
  \fslm = (-1)^\spin \: \img^{\m+\spin} \nl
  \summptrunc
  \dlmnhalfpim \:
  \dlmnhalfpisn \:
  \Gsmm
  \spcend ,
\end{equation}
where 
\begin{equation}
  \label{eqn:Gmm}
  \Gsmm = \intsaa \: \Gsmt \: \exp{-\img \m\p \saa}
\end{equation}
and
\begin{equation}
  \label{eqn:Gmt}
  \Gsmt = \intsab \: \fs(\sas) \: \exp{-\img \m \sab}
  \spcend .
\end{equation}
In \sectn{\ref{sec:fsht:forward}} we consider implicit quadrature rules to evaluate \eqn{\ref{eqn:Gmm}} and
\eqn{\ref{eqn:Gmt}} exactly.  By noting the same substitutions and interchanging the order of summation, the inverse transform of \fslm\ may be written
\begin{equation}
  \label{eqn:f_pfft}
  \fs(\sas) =  \summtrunc \: \Fsmt \: \exp{\img \m \sab}
  \spcend ,
\end{equation}
where
\begin{equation}
  \label{eqn:Fmt}
  \Fsmt = \summptrunc \: \Fsmm \: \exp{\img \m\p \saa}
\end{equation}
and
\begin{equation}
  \label{eqn:Fmm}
  \Fsmm  = (-1)^\spin \: \img^{-(\m+\spin)} 
  \sumltrunc \nl \:
  \dlmnhalfpim \:
  \dlmnhalfpisn \:
  \fslm
  \spcend .
\end{equation}
Although recasting the forward and inverse spherical harmonic transforms in this manner is no more efficient than the original formulation, expressions \eqn{\ref{eqn:Gmm}}--\eqn{\ref{eqn:Fmt}} highlight similarities with Fourier series representations.  However, the Fourier series expansion is only defined for periodic functions; thus, to recast these expressions in a form amenable to the application of Fourier transforms we must make a periodic extension in colatitude \saa.

%==============================================================================
\subsection{Periodic extension}

We make a periodic extension of \saa\ to the domain $[0,2\pi)$ so that \fft s may be used to compute the forward and inverse spherical harmonic transform rapidly.  When making this periodic extension we must be careful to ensure that the symmetry of our current representation is respected on the new domain; we must apply the symmetries dictated by the inverse transform when imposing periodisation in the forward transform.  By substituting \eqn{\ref{eqn:f_pfft}} into \eqn{\ref{eqn:Gmt}} and noting the continuous orthogonality of the complex exponentials, we find the forward and inverse expressions in \saa\ are related by 
%$\Gsmt = 2 \pi \Fsmt$.
\begin{equation}
  \label{eqn:Gmt_Fmt}
  \Gsmt %= \intsab \:  \summp \: \Fsmpt \: \exp{\img \m\p \sab} \: \exp{-\img \m \sab} 
  = 2 \pi \: \Fsmt
  \spcend .
\end{equation}
Consequently, the symmetry we impose in \Gsmt\ when extended periodically must match the symmetry of \Fsmt.  By reflecting \saa, we obtain the following symmetry for $\Fsmt$:
\begin{equation*}
 \Fsmtn 
 = \summp \: \Fsmm \: \exp{-\img \m\p \saa} 
 = (-1)^{\m+\spin} \: \Fsmt
 \spcend ,
\end{equation*}
where we have noted the symmetry 
\begin{equation}
  \label{eqn:Fmm_sym1}
  \Fsmmn = (-1)^{\m+\spin} \: \Fsmm
\end{equation}
following from \eqn{\ref{eqn:Fmm}} and \eqn{\ref{eqn:wignerd_sym2}}.  Thus, we extend \Gsmt\ to the $[0,2\pi)$ domain by constructing\footnote{We check that this periodic extension does not impose discontinuities at the poles \mbox{$\saa^\star \in \{ 0, \pi\}$}.  To avoid discontinuities we require \mbox{$\Fsm(\saa^\star) = (-1)^{\m+\spin} \Fsm(\saa^\star)$}, which follows trivially for \mbox{$\m+\spin$} even.  From \eqn{\ref{eqn:wignerd_sym3}} we find $\dmatsmall_{\m,-\spin}^\el(\saa^\star) = (-1)^{\m+\spin} \: \dmatsmall_{\m,-\spin}^\el(\saa^\star)$, which, due to the continuity of the Wigner \dmatsmall-functions, implies $\dmatsmall_{\m,-\spin}^\el(\saa^\star)=0$ for \mbox{$\m+\spin$} odd.
Combined with the representation, for \mbox{$\m+\spin$} odd,
\begin{equation*}
  \label{eqn:Fmt_alternative}
  \Fsmt 
  = (-1)^\spin \: \sumltrunc \: \nl \:  \dmatsmall_{\m,-\spin}^\el(\saa) \: \fslm
  \spcend ,
\end{equation*}
this implies $\Fsm(\saa^\star)=0$ for \mbox{$\m+\spin$} odd; hence our periodic extension does \emph{not} impose any discontinuity.}
\begin{equation*}
 \rGsmt = 
 \begin{cases}
   \: \Gsmt \: , & \saa \in [0,\pi]\\
   \: (-1)^{\m+\spin} \: \Gsm(2\pi - \saa) \: , & \saa \in (\pi,2\pi)
 \end{cases}
 \spcend .
\end{equation*}

Note that we adopt a different periodic extension to other approaches framed on the torus \cite{mcewen:fsht,huffenberger:2010} by applying the extension to the Fourier transform of \fs\ in \sab, \ie\ to \Gsmt.  Two periodic extensions, one even and one odd, were required in the approach taken previously by one of the authors \cite{mcewen:fsht}.  Huffenberger \& Wandelt \cite{huffenberger:2010} apply the $(-1)^\m$ factor as a shift in \sab\ by $\pi$, removing the need for two periodic extensions but requiring a even number of samples in \sab, precluding an association with the odd number of points in \m\ unless oversampling is performed.  We avoid these restrictions by applying the periodic extension to the Fourier transform in \sab\ of \fs, rather than to \fs\ directly.

%==============================================================================
\subsection{Discretisation}

We adopt an equiangular sampling of the sphere with sample positions given by 
\begin{equation}
\label{eqn:nodes_theta}
\saa_\saai = 
\frac{\pi(2\saai+1)}{2\elmax-1}, 
\quad \mbox{where } \saai \in \{ 0,1,\dotsc,\elmax-1 \}
\end{equation}
and
\begin{equation}
\label{eqn:nodes_phi}
\sab_\sabi = 
\frac{2 \pi \sabi}{2\elmax-1}, 
\quad \mbox{where } \sabi \in \{ 0,1,\dotsc,2\elmax-2 \}
\spcend .
\end{equation}
In order to extend the \saa\ domain to $[0, 2\pi)$ we simply extend the domain of the \saai\ index to include $\{ \elmax,\elmax+1,\dotsc,2\elmax-1 \}$.  

An odd number of sample points are required in both \saa\ and \sab\ in the extended domain so that a direct association may be made with the harmonic indices $\m$ and $\m\p$, resulting in \mbox{$\Nmw = (\elmax - 1) (2 \elmax - 1) + 1 \sim 2 \elmax^2$} samples on the sphere.  We also require a symmetric sampling in $\saa$ about the South pole so that samples on the extended domain can be obtained by reflecting samples defined on the original domain.  The node positions specified by \eqn{\ref{eqn:nodes_theta}} and \eqn{\ref{eqn:nodes_phi}} eliminate repeated samples at the poles $\saa=0$ and $\saa=2\pi$ since these points are excluded from the pixelisation.  However, it is not possible to eliminate repeated samples at $\saa=\pi$, since we require a discretisation that is symmetric about $\pi$ but which contains an odd number of sample points.

%==============================================================================
\subsection{Forward transform}
\label{sec:fsht:forward}

The algorithm we derive in this section to compute a forward spin spherical harmonic transform essentially follows the harmonic formulation presented in \sectn{\ref{sec:fsht:formulation}}, however we discuss implicit quadrature rules for the exact evaluation of \eqn{\ref{eqn:Gmm}} and \eqn{\ref{eqn:Gmt}}.  

Since \eqn{\ref{eqn:Gmt}} is simply a Fourier transform we may appeal to the discrete and continuous orthogonality of the complex exponential to express this integral exactly by
\begin{equation*}
 \Gsmti = \frac{2 \pi}{2\elmax-1} 
 \sumsabi \: 
 \fs(\saisang) \: 
 \exp{-\img \m \sabiang}
 \spcend ,
\end{equation*}
for $\saai \in \{ 0, 1, \dotsc, \elmax-1 \}$.  An \fft\ may be used to compute \Gsmti\ for all \m\ and \saai\ with computational complexity $\order(\elmax^2 \log_2 \elmax)$.  We extend \Gsmt\ to the domain $\saa \in [0,2\pi)$ through the construction 
\begin{align*}
 &\rGsmti = \\ 
 &\quad
 \begin{cases}
    \Gsmti \: , & \saai \in \{ 0, 1, \dotsc, \elmax-1 \} \\
   (-1)^{\m+\spin} \: \Gsm(\saa_{2\elmax-2-\saai}) \: , & \saai \in \{ \elmax, \dotsc, 2\elmax-2 \}
 \end{cases}
 \spcend ,
\end{align*}
noting $2\pi - \saaiang = \saa_{2\elmax-2-\saai}$.  

We now consider an implicit quadrature rule for the exact evaluation of \Gsmm\ through \eqn{\ref{eqn:Gmm}}.  Firstly, however, we compute \Fsmm\ from \rGsmti\ by noting \eqn{\ref{eqn:Gmt_Fmt}} and by appealing to the discrete orthogonality of the complex exponentials to invert \eqn{\ref{eqn:Fmt}}, giving
\begin{equation*}
 \Fsmm =
% \frac{(2 \pi)^{-1}}{2\elmax-1} 
 \frac{1}{2 \pi(2\elmax-1)} 
 \sumsaai \: \rGsmti \:
 \exp{- \img \m\p \saaiang}
 \spcend .
\end{equation*}
An \fft\ may be used to compute \Fsmm\ for all $\m$ and $\m\p$ with computational complexity $\order(\elmax^2 \log_2 \elmax)$.  Now that we have \Fsmm\ to hand, we substitute \eqn{\ref{eqn:Fmt}} into \eqn{\ref{eqn:Gmm}}, noting \eqn{\ref{eqn:Gmt_Fmt}}, to yield 
\begin{align}
 \Gsmm 
 &= \intsaa \: \rGsmt \: \exp{-\img \m\p \saa} \nonumber \\
 &= 2 \pi \sum_{\m{\p}{\p}=-(\elmax-1)}^{\elmax-1} \: \Fsmmp \: \weight(\m{\p}{\p} - \m\p)
 \spcend ,
 \label{eqn:Gmm_convolution}
\end{align}
where the weights are given by
\begin{align*}
 \weight(\m\p)
 &= \intsaa \: \exp{\img \m\p \saa} \\
 &= 
 \begin{cases}
   \: \pm \img \pi/2, & \m\p=\pm 1\\
   \: 0, & \m\p \text{ odd},\:\m\p\neq\pm1\\
   \: 2/(1-{\m\p}^2), & \m\p \text{ even}
 \end{cases}
 \spcend .
\end{align*}
Note that the definition of these weights is identical to that derived by Huffenberger \& Wandelt \cite{huffenberger:2010}, however we correct some (typographical) errors in their explicit evaluation.  
Since we are concerned with the values of \Gsmm\ for $|\m\p| \leq \elmax-1$, the computation of \Gsmm\ through \eqn{\ref{eqn:Gmm_convolution}} explicitly requires weights with argument up to $\pm 2(\elmax-1)$.  If the range of $\m\p$ is extended to $|\m\p| \leq 3(\elmax-1)$, \eqn{\ref{eqn:Gmm_convolution}} may be seen as a (reflected) convolution, which may be computed more efficiently following a Fourier transform.  However, since only the range $|\m\p| \leq \elmax-1$ is of interest, aliasing may be tolerated provided that it is outside of this range.  To ensure that this is the case we zero-pad \Fsmm\ in the domain $|\m\p| \in \{ \elmax, \dotsc, 2 \elmax - 2 \}$ prior to computing an inverse Fourier transform.  We then compute an inverse Fourier transform of the weights on the same extended domain and take the product of these terms, the Fourier transform of which gives \Gsmm.  Using \fft s to compute \Gsmm\ in this manner reduces the computational complexity from $\order(\elmax^3)$ for the direct calculation of \eqn{\ref{eqn:Gmm_convolution}} to $\order(\elmax^2 \log_2 \elmax)$.

Once we have computed \Gsmm\ through the implicit quadrature rule discussed previously, we simply compute the spin spherical harmonic coefficients \fslm\ through \eqn{\ref{eqn:flm_wig}}.  The complexity of this computation is $\order(\elmax^3)$, which dominates the overall complexity of the forward algorithm.  

The forward algorithm may be summarised conceptually as follows, where we view the upsampling and application of weights in the spatial domain: 
\begin{algorithmic}[1] \Procedure{Forward Transform}{ $\fs$ }
\State compute the Fourier transform of $\fs$ in \sab
\State extend the resultant function to $2\pi$ in \saa
\State upsample the resultant function in \saa 
\State multiply by the inverse Fourier transform of the 
\Statex \hspace{\algorithmicindent} reflected weights and take the Fourier transform \Statex \hspace{\algorithmicindent} in \saa\ to give the coefficients \Gsmm
\State compute the spherical harmonic coefficients $\fslm$ 
\Statex \hspace{\algorithmicindent} from \Gsmm
\State \textbf{return} $\fslm$ \EndProcedure
\end{algorithmic}

Although the complexity of this approach remains identical to a standard separation of variables, the continual use of \fft s reduces the constant prefactor associated with the asymptotic scaling considerably, resulting in an algorithm that may be used to compute harmonic coefficients rapidly.  Furthermore, a precomputation is not required, which can otherwise necessitate very large storage requirements for high band-limits.  Finally, note that this algorithm applies to both scalar and spin functions without any change in computational complexity or computation time.

%==============================================================================
\subsection{Inverse transform}

The inverse algorithm follows the harmonic formulation presented in \sectn{\ref{sec:fsht:formulation}} closely.  Firstly, \Fsmm\ is computed directly through \eqn{\ref{eqn:Fmm}}, where we also exploit the symmetry \eqn{\ref{eqn:Fmm_sym1}} to reduce the number of computations by a factor of two.  The complexity of this computation is $\order(\elmax^3)$, which dominates the overall complexity of the inverse algorithm. \fft s are then used to evaluate \eqn{\ref{eqn:Fmt}} and \eqn{\ref{eqn:f_pfft}} rapidly over the extended domain, with complexities $\order(\elmax ^2 \log_2 \elmax)$ for both computations.  To facilitate the efficient calculation of \eqn{\ref{eqn:f_pfft}} through the use of an \fft, we compute function values on the extended \saa\ domain $[0,2\pi)$, however we discard those values computed in the domain $(\pi,2\pi)$.  The algorithm presented here to compute the inverse transform is identical to that first proposed by one of the authors \cite{mcewen:fsht} and subsequently adopted by Huffenberger \& Wandelt \cite{huffenberger:2010}.
The inverse algorithm may be summarised as follows: 
\begin{algorithmic}[1] \Procedure{Inverse Transform}{ $\fslm$ }
\State compute the Fourier coefficients \Fsmm\ from \fslm
\State compute the function samples on the extended 
\Statex \hspace{\algorithmicindent} domain by an inverse Fourier transform
\State construct $\fs$ by discarding samples computed in the 
\Statex \hspace{\algorithmicindent} \saa\ domain $(\pi,2\pi)$ 
\State \textbf{return} $\fs$ \EndProcedure
\end{algorithmic}

Since we construct algorithms to perform forward and inverse spherical harmonic transforms that are theoretically exact, our construction corresponds to a novel sampling theorem on the sphere.  Moreover, to represent a band-limited signal on the sphere our sampling theorem requires less than half the number of number of samples required by other equiangular sampling theorems on the sphere \cite{shukowsky:1986,driscoll:1994,huffenberger:2010} and an asymptotically identical, but smaller, number of samples that the Gauss-Legendre sampling theorem.

%==============================================================================
\subsection{Quadrature}

\begin{figure}
\centering
\mbox{
\subfigure[Weights for $\elmax=4$]{\includegraphics[width=40mm]{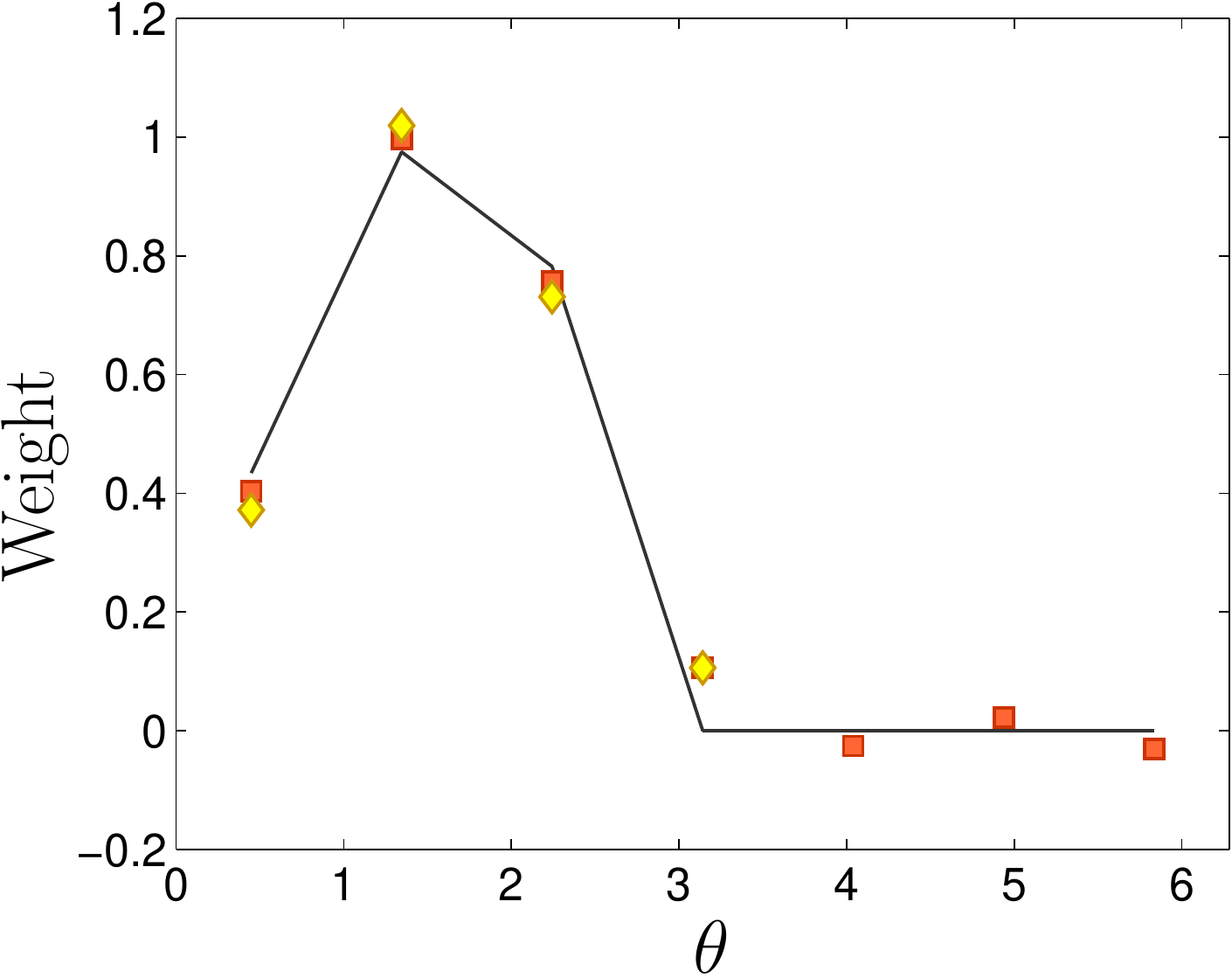}} \quad
\subfigure[Difference for $\elmax=4$]{\includegraphics[width=40mm]{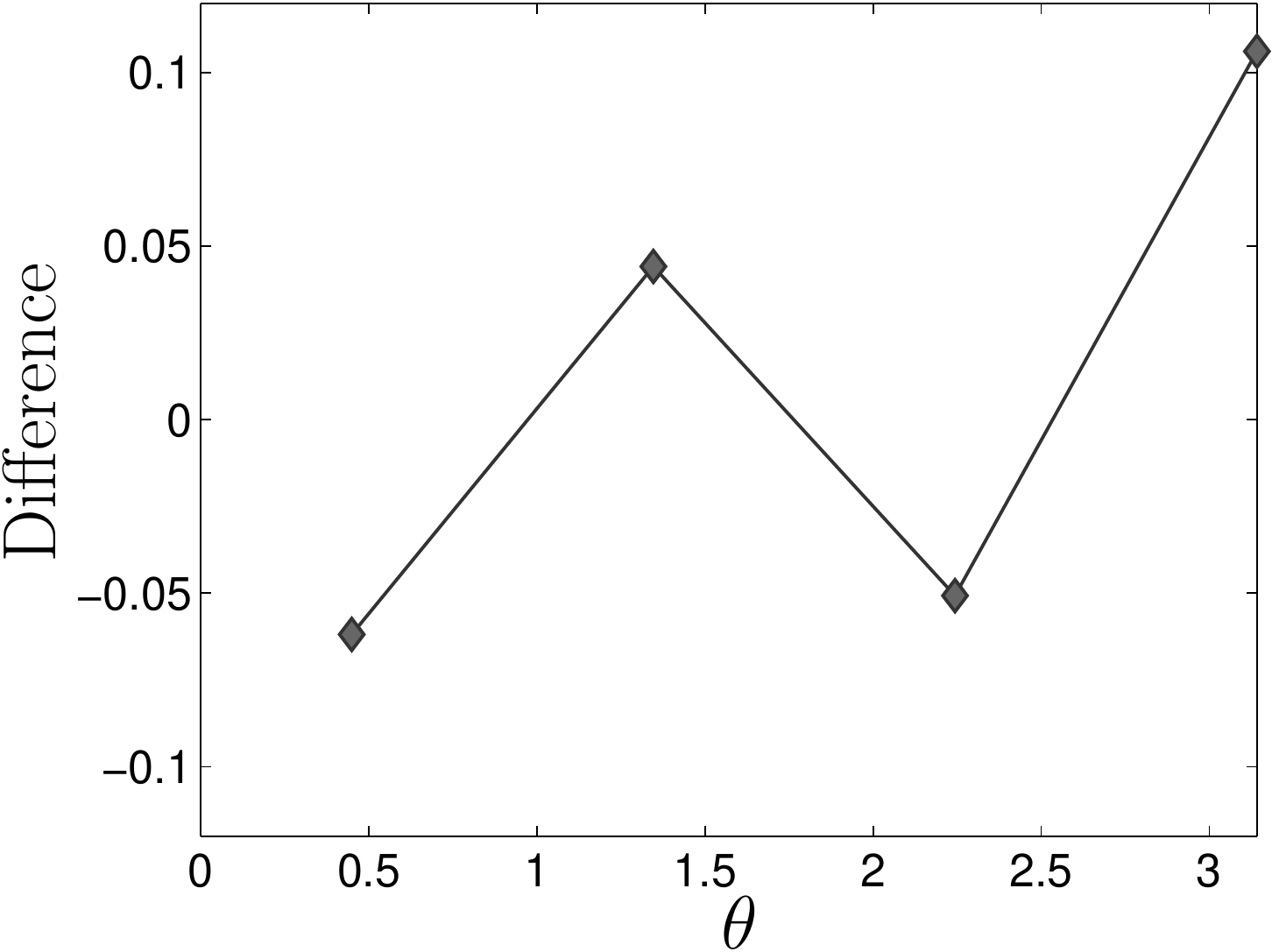}}
}
\mbox{
\subfigure[Weights for $\elmax=64$]{\includegraphics[width=40mm]{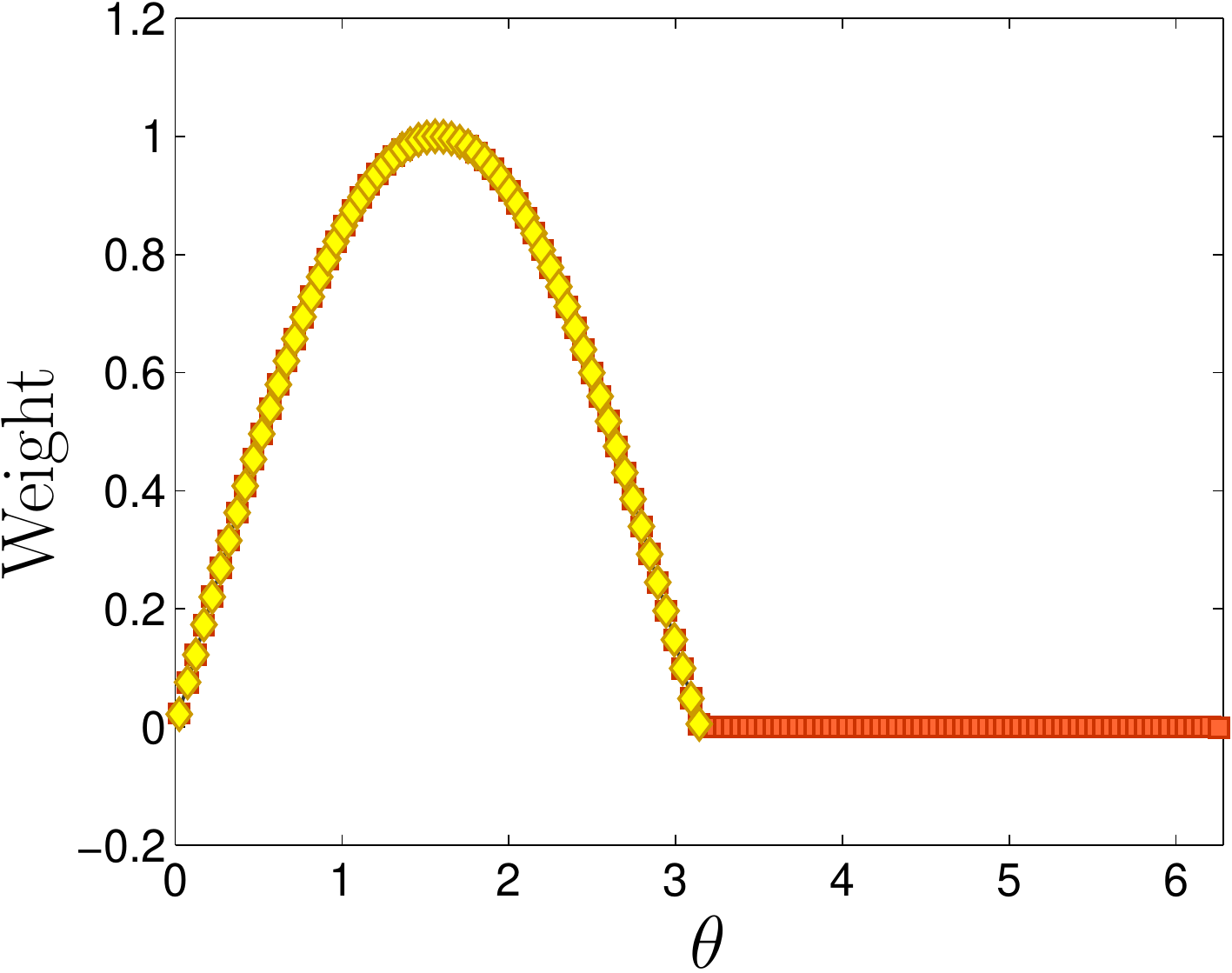}} \quad
\subfigure[Difference for $\elmax=64$]{\includegraphics[width=40mm]{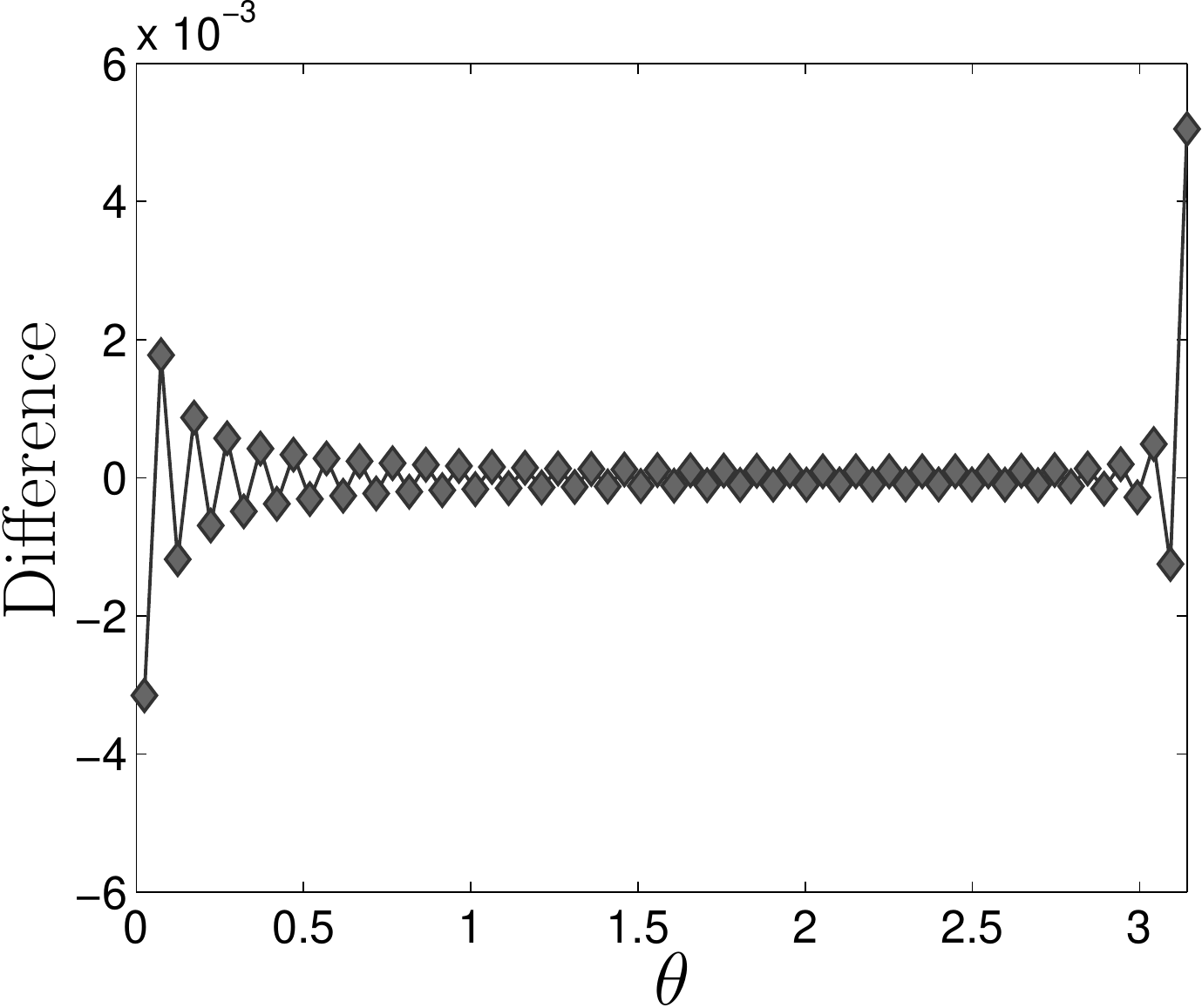}}
}
\caption{Exact quadrature weights corresponding to our sampling theorem. In the left column of panels the weights $\weighttrans(\saaiang)$ (red squares) defined on $[0, 2\pi)$ and the quadrature weights $\qweight(\saaiang)$ (yellow diamonds) defined on $[0, \pi]$ are plotted.  These values are compared to samples of the function defined by $\sin(\saa)$ on $[0, \pi)$ and zero on $[\pi, 2\pi)$ (solid black line).  In the right column of panels the difference between the quadrature weights  $\qweight(\saaiang)$ and $\sin(\saa)$ are plotted.}
\label{fig:weights}
\end{figure}

The construction of our sampling theorem on the sphere can be used to define an explicit quadrature rule for the integration of a function band-limited at $\elmax$.  This integration, which corresponds to computing the spherical harmonic coefficient $\shc{\fs}{0}{0}$, requires approximately half as many samples as needed to compute all spherical harmonic coefficients.  We define the explicit quadrature weights $\qweight(\saaiang)$ to evaluate the following integral exactly by the finite sum:
\begin{equation}
\label{eqn:quadrature}
I = \int_\sphere \dmu{\sas} \: \fs(\sas) = 
\sum_{\saai=0}^{\elmax-1} \:
\sum_{\sabi=0}^{\elmax-1} \:
\fs(\saisang\p) \: \qweight(\saaiang)
\spcend ,
\end{equation}
where $\sab_\sabi\p = 2 \pi \sabi / \elmax$.  As seen from \eqn{\ref{eqn:flm_wig}}, the computation of $\shc{\fs}{0}{0}$ requires $\Gsmm$ for $\m=\m\p=0$ only.  Consequently, aliasing in $\m$ in all Fourier coefficients \Gsmm\ except $\m=0$ may be tolerated, hence the number of samples required in \sab\ is reduced from $2\elmax-1$ to $\elmax$.  
From the (reflected) convolution \eqn{\ref{eqn:Gmm_convolution}}, it is apparent that the coefficients \Fsmm\ are required for all $\m\p$ (for $\m=0$ only).  Consequently, the sampling in \saa\ remains unchanged, with $\elmax$ samples.  However, only weights with argument up to $\pm (\elmax-1)$ are required.  The (reflected) convolution thus spans the range $|\m\p| \leq 2(\elmax-1)$.  However, since only $\m\p=0$ is of interest aliasing may be tolerated in $\m\p$ in all Fourier coefficients \Gsmm\ except $\m\p=0$, so that zero-padding is not required before computing the (reflected) convolution as a product in the spatial domain.  
The absence of upsampling leads to the explicit quadrature \eqn{\ref{eqn:quadrature}}, with \mbox{$\elmax(\elmax-1)+1$} samples on the sphere, where the weights are defined by
\begin{equation*}
\qweight(\saaiang) = 
\frac{2\pi}{\elmax} 
\Bigl[ 
\weighttrans(\saaiang) 
+ (1 - \kron{t,}{\elmax-1}) \: (-1)^\spin \: \weighttrans(\saa_{2\elmax-2-\saai})
\Bigr]
\spcend ,
\end{equation*}
and where $\weighttrans(\saaiang)$ is the inverse discrete Fourier transform of the reflected weights $\weight(-\m\p)$:
\begin{equation*}
\weighttrans(\saaiang) =
\frac{1}{2\elmax-1} \: \sum_{\m\p=-(\elmax-1)}^{\elmax-1} \:
\weight(-\m\p) \: \exp{\img \m\p \saaiang}
\spcend .
\end{equation*}
The weights $\weighttrans(\saaiang)$ defined on $[0, 2\pi)$ are exactly the samples of the function defined by $\sin(\saa)$ on $[0, \pi)$ and zero on $[\pi, 2\pi)$, band-limited at \elmax.  The quadrature weights $\qweight(\saaiang)$ defined on $[0, \pi]$ are constructed simply by folding the contributions of $\weighttrans(\saaiang)$ on $(\pi, 2\pi)$ back onto the $[0, \pi]$ domain.  Both of these weights are plotted and compared to $\sin(\theta)$ in \fig{\ref{fig:weights}}.

%==============================================================================
\subsection{Real signals}

In many practical applications signals observed on the sphere satisfy a reality condition.  For spin signals, the reality condition is given by \mbox{$\fs^\cconj=\fsm$}, which implies the conjugate symmetry condition $\shcc{\fs}{\el}{\m} = (-1)^{\spin+\m} \: {}_{-\spin}\shcsp{\f}{\el}{-\m}$ on the spherical harmonic coefficients of the signal.  When this reality condition is satisfied (for example, when considering the polarisation of the \cmb), we may exploit this symmetry to recover the harmonic coefficients of a spin $-\spin$ signal for free from the coefficients of a spin $\spin$ signal.  For the spin $\spin=0$ case, the reality condition reduces to the standard reality condition of a scalar signal, implying \mbox{$\shcc{f}{\el}{\m} = (-1)^\m \: \shcsp{f}{\el}{-\m}$}.  In this case, noting \eqn{\ref{eqn:Fmm}} and \eqn{\ref{eqn:wignerd_sym1}}, we obtain the symmetry
\begin{equation*}
%  \label{eqn:Fmm_sym2}
  \Fzmnmn = \Fzmm^\cconj
  \spcend .
\end{equation*}
We exploit these symmetries to reduce the computational cost of both the forward and inverse algorithms by an additional factor of approximately two for real spin $\spin=0$ signals.

%==============================================================================
\section{Evaluation}
\label{sec:evaluation}
%==============================================================================

We have implemented our fast algorithms to compute spherical harmonic transforms in double precision arithmetic, exploiting all of the symmetries discussed in \sectn{\ref{sec:fsht}} to optimise the implementation.\footnote{The structure of our algorithms suggest multiple transforms of different spin may in theory be computed simultaneously at lower cost than consecutive computation  (since Wigner \dmatsmall-functions do not need to be recomputed and some computations are independent of spin).  However, for simplicity we do not incorporate this optimisation in our current implementation.}  The core implementation used for the numerical experiments presented in this section is written in {\tt C}, using the {\tt FFTW}\footnote{\url{http://www.fftw.org/}} package to compute Fourier transforms, however we also provide a {\tt MATLAB} interface.  We make our Spin Spherical Harmonic Transform ({\tt SSHT}) package containing this implementation available publicly.\footnote{\url{http://www.jasonmcewen.org/}}  
For comparison purposes, we also implemented in the {\tt SSHT} package an optimised algorithm to compute spherical harmonic transforms for the Gauss-Legendre sampling theorem on the sphere.\footnote{It is well know that Gauss-Legendre quadrature may be used to construct an exact sampling theorem on the sphere.  Gauss-Legendre quadrature with $P$ points is exact only for a polynomial integrand of order less than or equal to $2P-1$.  Since neither the associated Legendre functions nor the Wigner \dmatsmall-functions are polynomials in $\cos\saa$, it does \emph{not} follow immediately that Gauss-Legendre quadrature results in an exact harmonic transform for scalar and spin signals on the sphere band-limited at \elmax; nevertheless, this is indeed the case.  We prove the result for spin \spin\ signals on the sphere, thus the scalar case will follow simply by setting $\spin=0$.  For \elmax\ samples in \saa, we must simply prove that the integrand $\dmatsmall_{\m\spin}^{\el}(\saa) \: \G{-\spin}{\m}{}(\saa)$ is polynomial in $\cos\saa$ of maximum degree less than or equal $2\elmax-1$.  From inspection of \eqn{\ref{eqn:wignerd_b}}, we may write the Wigner \dmatsmall-functions as a polynomial of degree $\el-\spin$, multiplied by $ ( {1-\cos\saa} )^{(\spin-\m)/2} ( {1+\cos\saa} )^{(\spin+\m)/2}$.  Similarly, we may write $\G{-\spin}{\m}{}(\saa)$ as a polynomial of degree $\elmax-1-\spin$, multiplied by the same factor.  The integrand is therefore polynomial with overall degree $\elmax-1+\el$, which reaches a maximum of $2\elmax - 2$.  Gauss-Legendre quadrature with $\elmax$ samples in \saa\ may thus be used to compute exact spherical harmonic transforms of scalar and spin functions band-limited at \elmax.}  This algorithm is based on a separation of variables and a direct application of the Gauss-Legendre quadrature rule, resulting in complexity $\order(\elmax^3)$.

In this section we evaluate our sampling theorem and fast algorithms in terms of the number of samples required to represent a band-limited function, numerical precomputation (or lack thereof), the recursions used to compute Wigner \dmatsmall-functions, and numerical accuracy and computation time.  Finally, we evaluate our sampling theorem in the context of potential applications.

%==============================================================================
\subsection{Sampling}

\begin{figure}
\centering
\includegraphics[width=80mm]{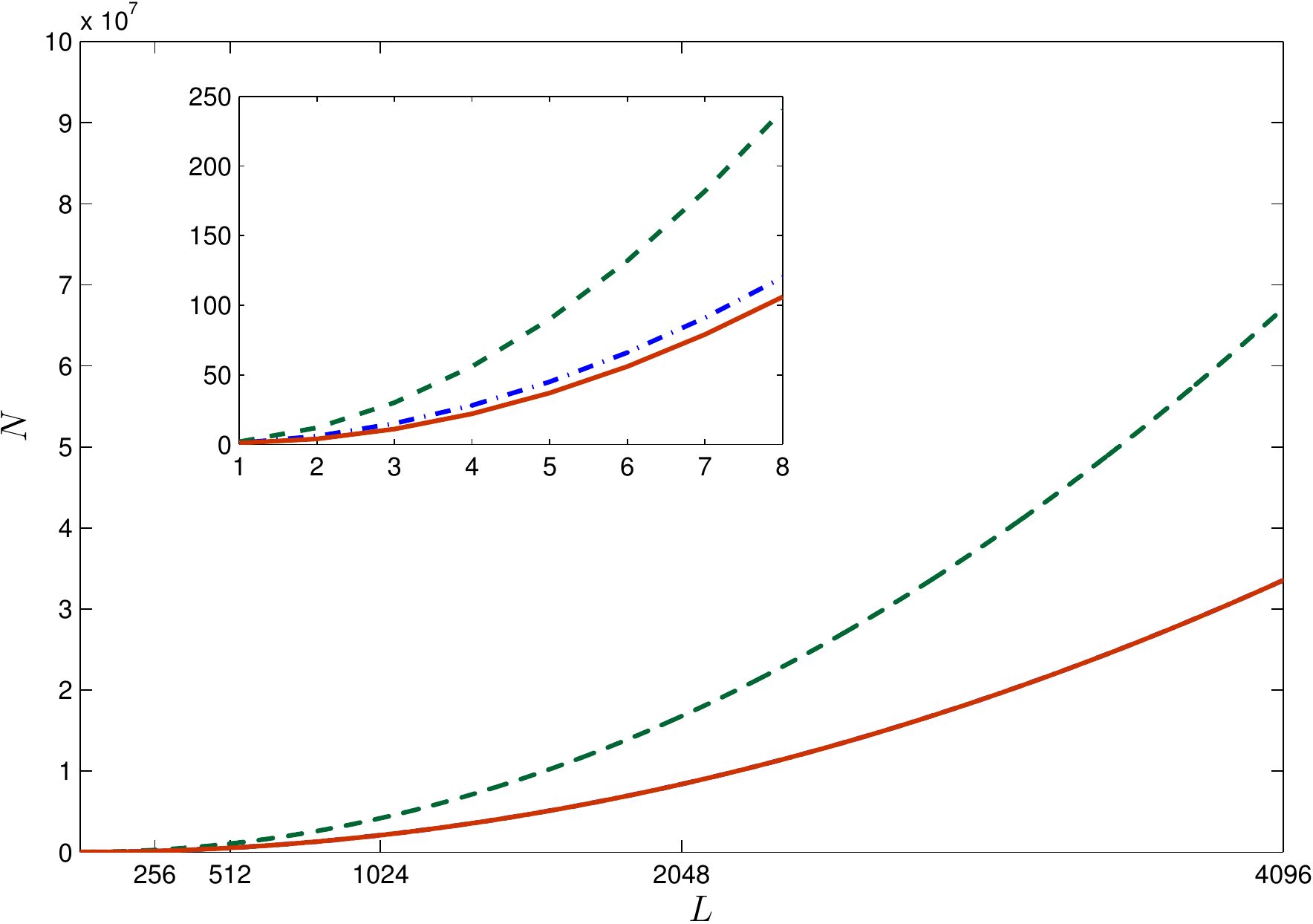}
\caption{Number of samples \N\ required to represent exactly a signal on the sphere of band-limit \elmax\ for the following sampling theorems: Gauss-Legendre sampling theorem (blue/dot-dashed line); Driscoll \& Healy sampling theorem (green/dashed line); and the sampling theorem developed in this article (red/solid line).  The inset shows very low band-limits, where the difference between Gauss-Legendre sampling and our sampling can have a large impact.}
\label{fig:n_vs_el}
\end{figure}

\begin{figure}
\centering
\subfigure[View of North pole]{\includegraphics[clip=,viewport=90 45 410 375,width=50mm]{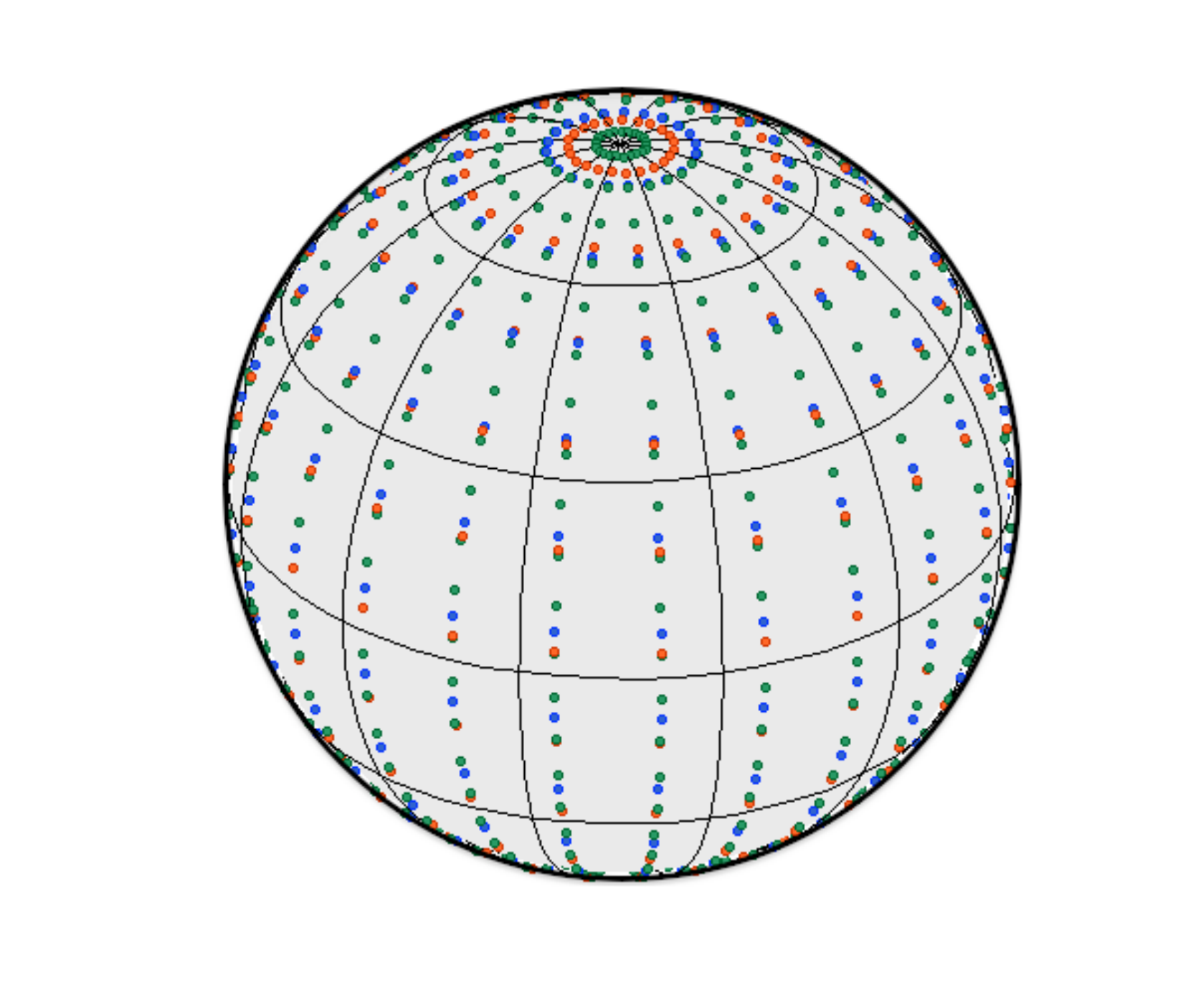}}
\subfigure[View of South pole]{\includegraphics[clip=,viewport=90 45 410 375,width=50mm]{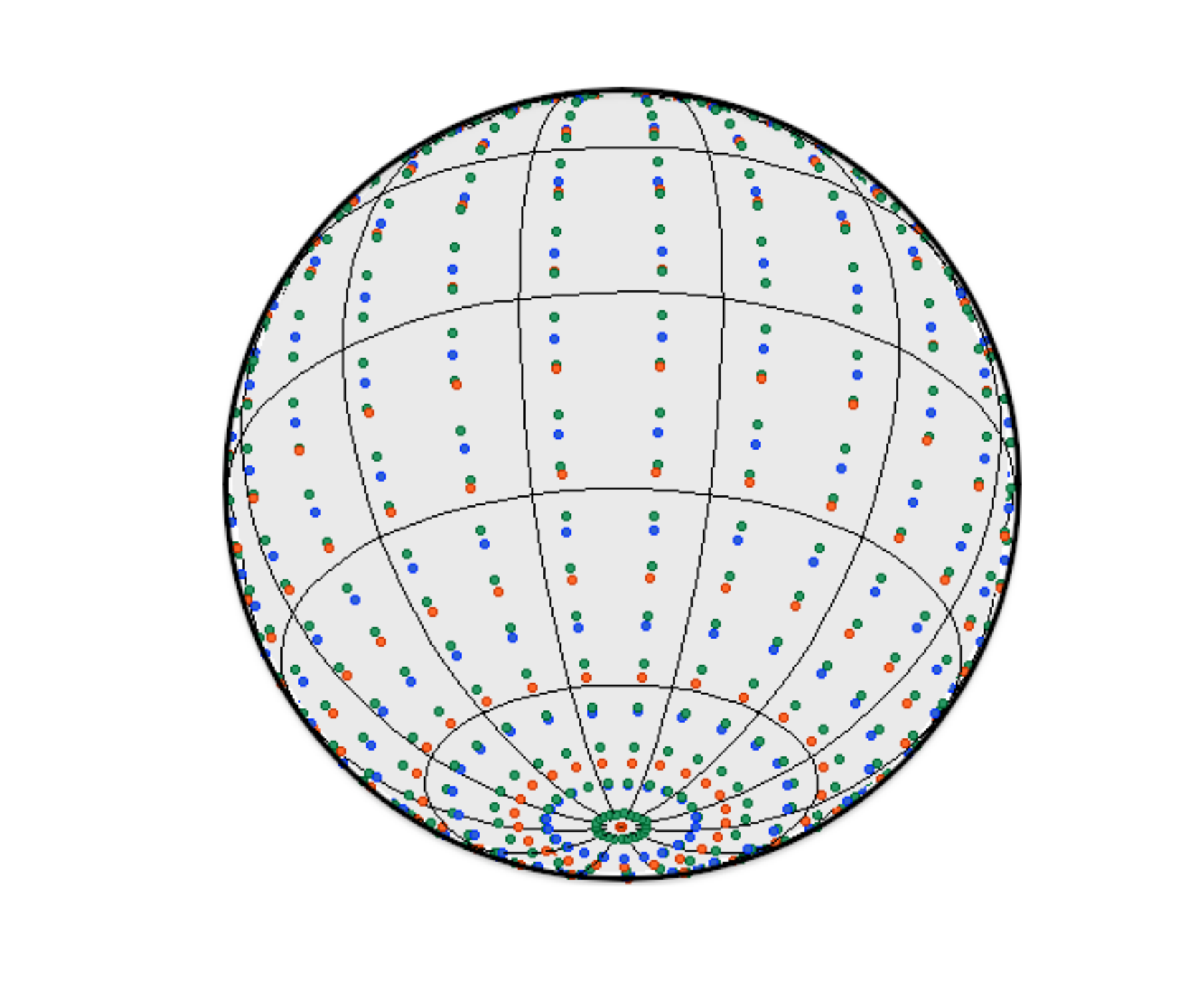}}
\caption{Sampling schemes for the exact representation of a signal band-limited at $\elmax=12$.  Sample positions are shown for the following sampling theorems: Gauss-Legendre sampling theorem (blue dots); Driscoll \& Healy sampling theorem (green dots); and the sampling theorem developed in this article (red dots).  Notice that the Driscoll \& Healy sampling theorem requires approximately twice as many samples on the sphere as the alternative samplings.}
\label{fig:samples}
\end{figure}

The number of samples required to represent a band-limited function on the sphere exactly is the fundamental property of any sampling theorem, with fewer samples desired.  Both the Gauss-Legendre and our sampling theorems require in general $\Ngl\sim\Nmw\sim 2\elmax^2$ samples, while the Driscoll \& Healy sampling theorem requires $\Ndh\sim 4 \elmax^2$ samples.  We therefore provide a reduction in number of samples by a factor of two compared to the canonical equiangular sampling theorem on the sphere.  Furthermore, we require \mbox{$\Ngl - \Nmw = 3(\elmax - 1)$} fewer samples that the Gauss-Legendre sampling theorem, which for small band-limits can be significant (as discussed in \sectn{\ref{sec:evaluation:applications}}). 
The optimal number of samples attainable by a sampling theorem on the sphere is given by the $\elmax^2$ degrees of freedom in harmonic space.  No sampling theorem on the sphere reaches this bound in general. However, both the Gauss-Legendre and our sampling theorem reach the $\elmax^2$ bound in the limiting case of small \elmax\ (we reach the bound for the cases $\elmax\in\{1,2\}$, the Gauss-Legendre sampling theorem reaches the bound for $\elmax=1$ only, while the Driscoll \& Healy sampling theorem never reaches the bound).
In \fig{\ref{fig:n_vs_el}} we plot the number of samples against band-limit for various sampling theorems.  We also plot the position of samples for each of these sampling theorems in \fig{\ref{fig:samples}}.

%==============================================================================
\subsection{Precomputation}

It is possible to reduce the computational burden of our fast algorithms by precomputing the Wigner \dmatsmall-functions for argument $\pi/2$ and for all required harmonic indices.  Such a precomputation would require $\order(\elmax^3)$ storage, similar to the storage requirements of the precomputation for Driscoll \& Healy based algorithms \cite{driscoll:1994,healy:2003,wiaux:2005b}.  For these latter approaches precomputation is essential to recover the fast algorithms with complexity below $\order(\elmax^3)$.  However, for high band-limits the storage requirements become impractical at present (recall that the precomputation at $\elmax=4096$ is expected to require approximately 77GB of storage).  The Wigner \mbox{\dmatsmall-functions} may be evaluated accurately and rapidly using recursion formulae; therefore to avoid storage problems we do not perform any precomputation and instead compute Wigner \dmatsmall-functions on-the-fly using the method of Risbo \cite{risbo:1996}.

%==============================================================================
\subsection{Computing Wigner functions}
\label{sec:evaluation:recursion}

Since our algorithms require Wigner \dmatsmall-functions evaluated on the entire $(\m,\m\p)$ plane for argument $\pi/2$ only, they are flexible with regard to the choice of recursion used to compute the Wigner plane for a given $\el$.  For example, for the on-the-fly computation of Wigner \dmatsmall-functions we may use the recursion of Risbo \cite{risbo:1996} (which requires the entire plane) or Trapani \& Navaza \cite{trapani:2006} (which is restricted to argument $\pi/2$), which are both of complexity $\order(\elmax^2)$, without altering the overall $\order(\elmax^3)$ complexity of our algorithms.  However, this is not the case for alternative algorithms.

To compute spherical harmonic transforms for the Gauss-Legendre and Driscoll \& Healy sampling theorems using the $\order(\elmax^3)$ algorithms described previously, it is necessary to compute Wigner \dmatsmall-function values for a single row of the Wigner plane only, but for all \el\ and \emph{all values} of \saa.  For the overall algorithms to remain $\order(\elmax^3)$, the on-the-fly computation of the row of the Wigner plane must be performed in $\order(\elmax)$ computations.  This precludes the use of the recursions devised by Risbo \cite{risbo:1996} and by Trapani \& Navaza \cite{trapani:2006}, for example, both of which would result in an overall algorithm with complexity $\order(\elmax^4)$.  Instead, alternative recursions must be used, such as the three-term recursion in \el\ that goes pointwise through the Wigner plane (see \eg\ (4.5) of \cite{kostelec:2008}; this recursion is used in our implementation of the Gauss-Legendre sampling theorem).  The inflexibility of these algorithms with regard to the choice of recursion used to compute Wigner \dmatsmall-functions becomes important when we study the stability of these recursions in the following section.  Of course, this issue may be resolved by precomputing Wigner \dmatsmall-functions using any recursion, but as we have seen this becomes problematic at large band-limits.

%==============================================================================
\subsection{Numerical accuracy and computation time}

We evaluate the numerical accuracy and computation time of our algorithms that implement our new sampling theorem, comparing them to our optimised implementation of the Gauss-Legendre sampling theorem and to the semi-naive algorithm \cite{healy:2003} in {\tt SpharmonicKit}\footnote{\url{http://www.cs.dartmouth.edu/~geelong/sphere/}} implementing the Driscoll \& Healy sampling theorem.  For all cases we do \emph{not} perform any precomputation since this is infeasible for high band-limits (recall that the semi-naive algorithm is the fastest algorithm implementing the Driscoll \& Healy sampling theorem that does not require precomputation).  In order to assess numerical accuracy and computation time we perform the following numerical experiment.  We generate band-limited test signals on the sphere defined by uniformly random spherical harmonic coefficients with real and imaginary parts distributed in the interval $[-1,1]$.  An inverse transform is performed to synthesise the test signal on the sphere from its spherical harmonic coefficients, followed by a forward transform to recompute harmonic coefficients.  Numerical accuracy is measured by the maximum absolute error between the original spherical harmonic coefficients $\fslm^{\rm o}$ and the recomputed values $\fslm^{\rm r}$, \ie\
%\begin{equation*}
$
  \epsilon = \mathop{\rm max}_{\el,\m} \:
  \bigl | \fslm^{\rm r} - \fslm^{\rm o} \bigr |
$.
%  \spcend .
%\end{equation*}
Computation time is measured by the round-trip computation time taken to perform the inverse and forward transform.  All numerical experiments are performed on a 2.5GHz Intel Pentium dual core processor with 4GB of RAM and are averaged over five random test signals.  

\begin{figure}
\centering
\includegraphics[width=80mm]{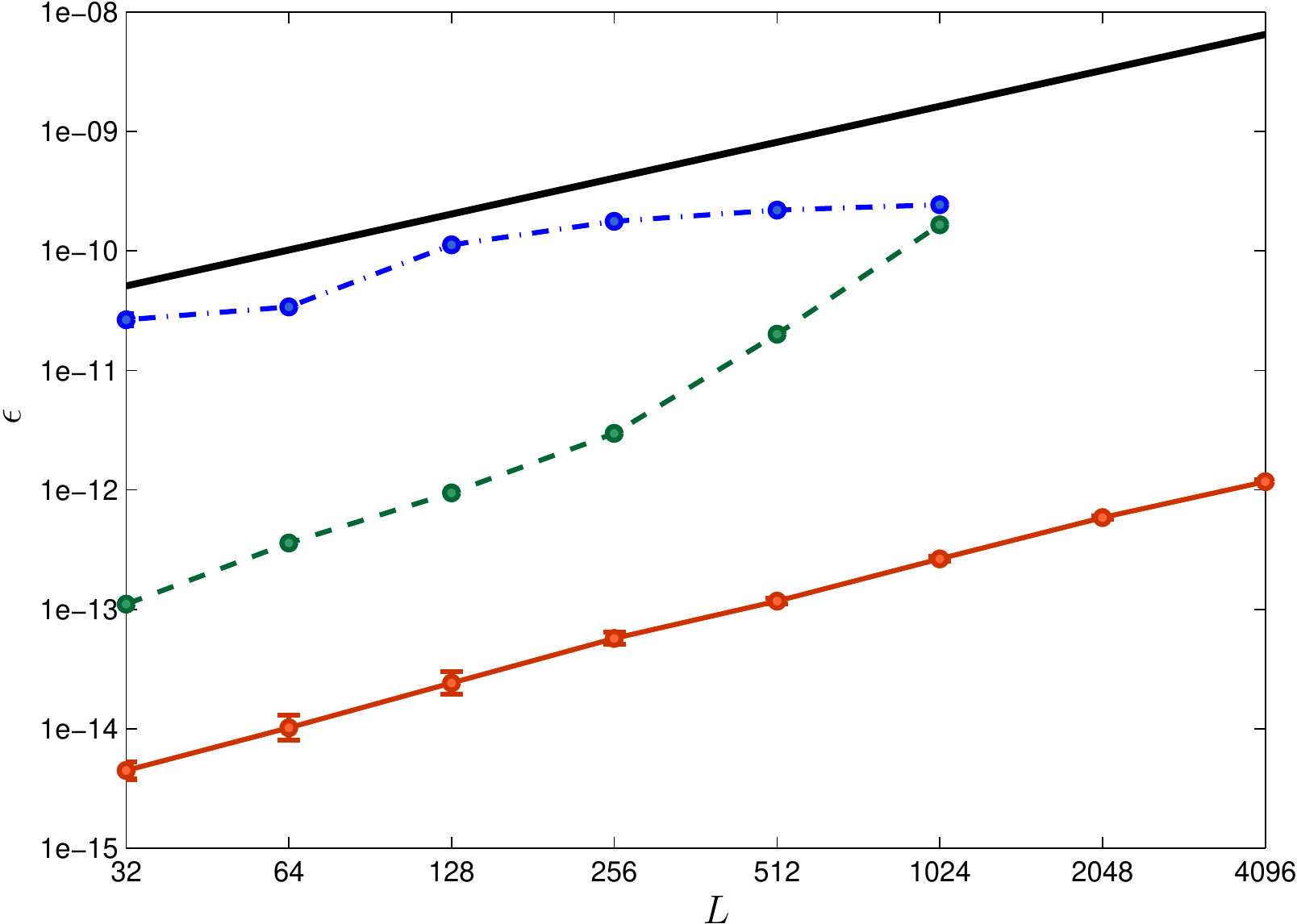}
\caption{Numerical accuracy of the algorithms implementing the following sampling theorems: our optimised implementation of the Gauss-Legendre sampling theorem (blue/dot-dashed line); the semi-naive algorithm in {\tt SpharmonicKit} implementing the Driscoll \& Healy sampling theorem (green/dashed line); and our algorithms implementing the sampling theorem developed in this article (red/solid line).  $\order(\elmax)$ scaling is shown by the heavy black/solid line.  
The algorithms implementing the Gauss-Legendre and Driscoll \& Healy sampling theorems go unstable between $\elmax=1024$ and $\elmax=2048$, due to the enforced use of the pointwise three-term Wigner recursion.
For the Gauss-Legendre and our sampling theorems, which both support spin transforms, the maximum absolute error $\epsilon$ is averaged over complex signals of spin $\spin\in\{0,2,10\}$ and a real spin $\spin=0$ signal, with one standard deviation error bars shown (in most cases differences are very small and error bars cannot be seen easily).  Note that for these cases the maximum absolute error is identical (to statistical noise) for transforms of real and complex signals of different spin.}
\label{fig:error}
\end{figure}

The maximum absolute error is plotted against band-limit in \fig{\ref{fig:error}} for different sampling theorems.  High numerical accuracy is achieved for all sampling theorems at moderate band-limits, with errors on the order of the machine precision and increasing approximately linearly with band-limit.  
For the Gauss-Legendre sampling theorem we use the {\tt gauleg} routine of Numerical Recipes \cite{press:1992} to compute Gauss-Legendre node positions and weights.  This method is based on an initial approximation for each node position, followed by an iterative refinement based on Newton's method, which is likely to explain the slightly inferior error performance of the corresponding algorithms.  Furthermore, the algorithms implementing both the Gauss-Legendre and Driscoll \& Healy sampling theorems suffer from their lack of flexibility regarding the recursion used to compute Wigner \dmatsmall-functions (or associated Legendre functions for the spin $\spin=0$ case), which necessitates the use of the less accurate pointwise three-term recursion in \el, rather than more accurate alternatives.  Consequently, the numerical accuracy of our algorithms is superior to the optimised implementations of the alternative sampling theorems.
More importantly, however, both the Gauss-Legendre sampling theorem and the semi-naive implementation of the Driscoll \& Healy sampling theorem go unstable between $\elmax=1024$ and $\elmax=2048$, due to the instability of the pointwise three-term Wigner recursion.  As discussed in \sectn{\ref{sec:evaluation:recursion}}, these algorithms require a recursion with complexity \order(\elmax), in order to remain $\order(\elmax^3)$ for on-the-fly computation.  To resolve the instability of these algorithms it would be necessary to use the alternative recursion of Risbo \cite{risbo:1996}, making them $\order(\elmax^4)$, or to perform a precomputation of Wigner \dmatsmall-functions.  Neither of these solutions are feasible for high band-limits, hence these algorithms are restricted to moderate band-limits (unless an alternative pointwise recursion can be found that is stable to high band-limits).  Due to the flexibility of our sampling theorem with regard to Wigner recursion, we are able to use Risbo's recursion \cite{risbo:1996}, which is stable to at least $\elmax=4096$, without altering the complexity of our algorithms.
Finally, note that for the implementations that support spin transforms (the Gauss-Legendre and our sampling theorems), the error is identical (to statistical noise) for transforms of real and complex signals of different spin.

\begin{figure}
\centering
\includegraphics[width=80mm]{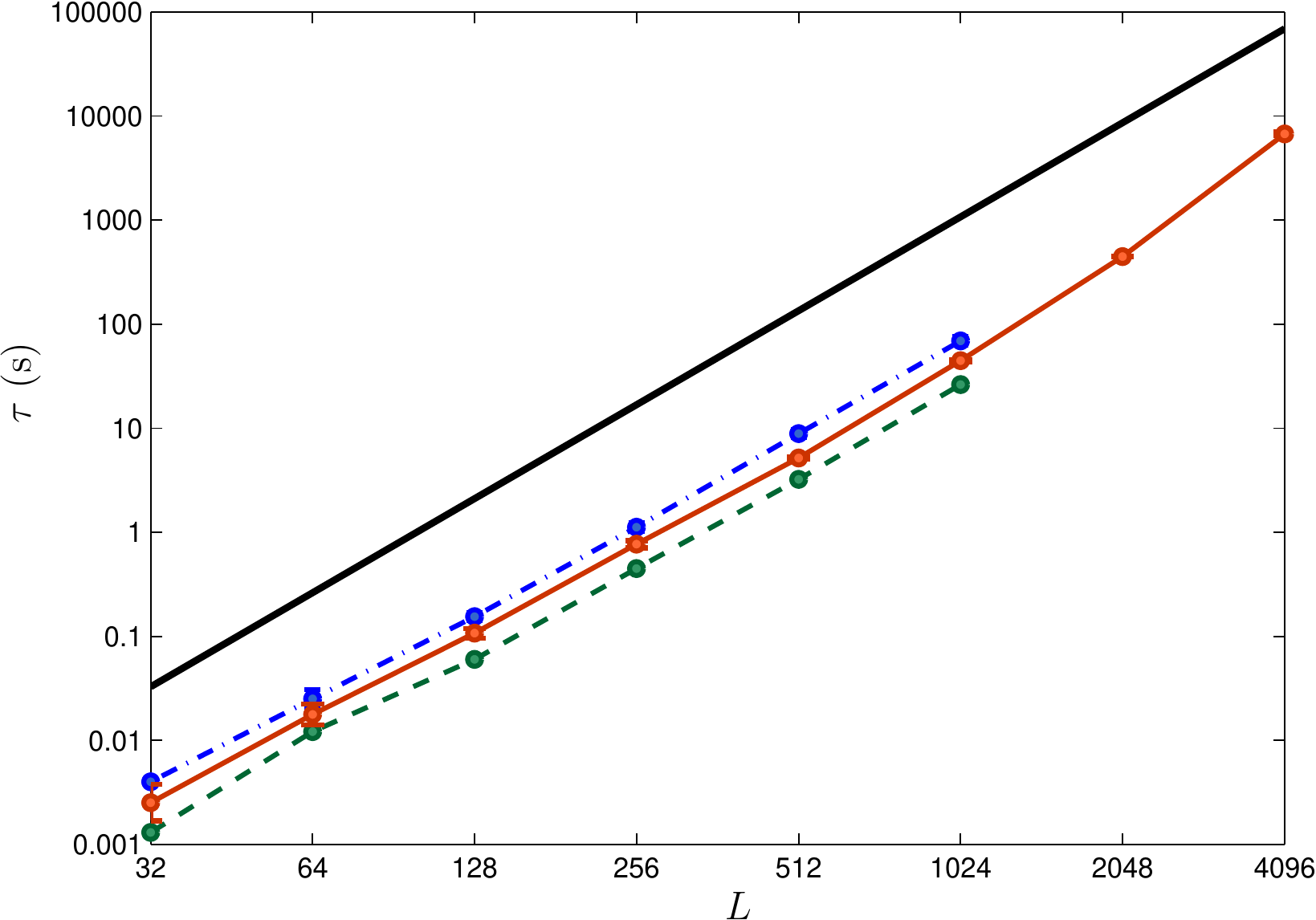}
\caption{Computation time of the algorithms implementing the following sampling theorems: our optimised implementation of the Gauss-Legendre sampling theorem (blue/dot-dashed line); the semi-naive algorithm in {\tt SpharmonicKit} implementing the Driscoll \& Healy sampling theorem (green/dashed line); and our algorithms implementing the sampling theorem developed in this article (red/solid line).  $\order(\elmax^3)$ scaling is shown by the heavy black/solid line. 
The algorithms implementing the Gauss-Legendre and Driscoll \& Healy sampling theorems go unstable between $\elmax=1024$ and $\elmax=2048$, due to the enforced use of the pointwise three-term Wigner recursion.
For the Gauss-Legendre and our sampling theorems, which both support spin transforms, the computation time $\tau$ (seconds) is averaged over complex signals of spin $\spin\in\{0,2,10\}$, with one standard deviation error bars shown (in most cases differences are very small and error bars cannot be seen easily).  Note that for these cases the computation time is identical (to statistical noise) for transforms of signals of different spin.}
\label{fig:timing}
\end{figure}

The computation time for complex signals is plotted against band-limit in \fig{\ref{fig:timing}} for different sampling theorems.  For all sampling theorems, computation time evolves as $\order(\elmax^3)$ as predicted.  The semi-naive algorithm is slightly faster than our algorithm, which is in turn slightly faster than our optimised implementation of the Gauss-Legendre sampling theorem.  Although we plot performance results for our algorithms using the recursion of Risbo \cite{risbo:1996}, we also implemented the recursion of Trapani \& Navaza \cite{trapani:2006}, which we found to be approximately 20\% faster than Risbo's approach but which goes unstable between $\elmax=2048$ and $\elmax=4096$.  Nevertheless, the Trapani \& Navaza \cite{trapani:2006} recursion can be used at band-limits at or below $\elmax=2048$ to provide a considerable speed enhancement.  In this case, at band-limit $\elmax=1024$ we are approximately 25\% slower than the semi-naive algorithm, but twice as fast as the Gauss-Legendre algorithm.  However, the semi-naive algorithm applies for scalar functions only (and requires approximately twice as many samples on the sphere), while the alternative sampling theorems also apply directly for spin functions on the sphere.  
%Furthermore, let us re-emphasise that the algorithms implementing our sampling theorem are the only ones feasible for high band-limits.
%
Finally, note that for the implementations that support spin transforms (the Gauss-Legendre and our sampling theorems), computation time is identical (to statistical noise) for transforms of signals of different spin, as predicted.

%==============================================================================
\subsection{Applications}
\label{sec:evaluation:applications}

We discuss three potential applications of our sampling theorem in the fields of cosmology, neuroscience and compressive sampling (CS).  For each case we highlight the enhancements that our sampling theorem will afford.

For the analysis of CMB observations, our sampling theorem and associated algorithms provide the ability to perform fast spherical harmonic transforms that are exact at the very high resolution of current and forthcoming CMB observations.  Furthermore, we can compute harmonic transforms of both the temperature and polarisation of the CMB for identical cost.

The number of samples required to represent a band-limited signal is not only of theoretical interest but also has important practical application.  
A sampling theorem requiring fewer samples means a band-limited function can be measured exactly for lower cost.  This is particularly important in applications where the cost of acquiring a single sample is large.  In neuroscience, for example, diffusion magnetic resonance imaging (MRI) \cite{johansenberg:2009} is one such application, where cost is measured in terms of acquisition time.  Diffusion MRI has received considerable attention recently as a non-invasive technique to image structural neuronal connectivity in the brain.  In this setting, most recent acquisition strategies consider sampling on multiple spherical shells for each voxel of the brain, from which an orientation distribution function (ODF) describing the probability density of neuronal fibre directions is recovered \cite{tuch:2004,canales:2009}.  The ODFs of each voxel are then combined to recover neuronal connectivity.  Given the millions of voxels generally considered, at present this imaging modality remains too time consuming for clinical use.  Typically, very low band-limits of order $\elmax\sim10$ are considered for each spherical shell.  Consequently, when adopting an exact sampling theorem even the small reduction in number of samples between our sampling theorem and the Gauss-Legendre approach of \mbox{$\Ngl - \Nmw = 3(\elmax - 1)$} can have a large impact on the total cost of acquisition.  For a typical example with acquisitions made on three concentric spherical shells of increasing radius, it has been shown that band-limits of three, five and nine, respectively, are required to limit aliasing to acceptable levels \cite{daducca:ssdmri}.  For such an acquisition, the total number of samples, and thus total acquisition time, would be reduced by a factor of 13\% when replacing Gauss-Legendre sampling with our sampling theorem.  This type of enhancement is of considerable importance in order to make diffusion MRI accessible for clinical use.

The recently developed theory of CS states that it is possible to acquire sparse or compressible signals with fewer samples than standard sampling theorems would suggest \cite{candes:2006a,donoho:2006}.  In these settings, the ratio of the number of required measurements to the dimensionality of the signal scales linearly with its sparsity \cite{candes:2006a}.  By reducing the dimensionality of the signal in the spatial domain, our sampling theorem will enhance the performance of CS reconstruction on the sphere when compared to alternative sampling theorems.  Furthermore, for sparsity priors defined in the spatial domain, such as signals sparse in the magnitude of their gradient, sparsity is also directly related to the sampling of the signal.  For this class of signals, we therefore expect to see an additional enhancement in CS reconstruction performance when adopting our sampling theorem.  The use of CS techniques on the sphere is likely to have wide-spread application for a wide range of problems, including more efficient acquisition, denoising and deconvolution on the sphere.  In particular, all of these problems are faced in diffusion MRI and in analysing the CMB, which we are studying currently to evaluate in detail the enhancements provided by our sampling theorem.

%==============================================================================
\section{Conclusions}
\label{sec:conclusions}
%==============================================================================

We have developed a novel sampling theorem on the sphere, with corresponding fast algorithms, by associating the sphere with the torus through a periodic extension.  To represent a band-limited signal on the sphere exactly our sampling theorem requires less than half the number of samples required by other equiangular sampling theorems on the sphere and an asymptotically identical, but smaller, number of samples than the Gauss-Legendre sampling theorem on the sphere.  The complexity of our algorithms to compute both forward and inverse transforms is $\order(\elmax^3)$, with identical scaling to a standard separation of variables.  However, the continual use of \fft s reduces the constant prefactor associated with the asymptotic scaling considerably, resulting in algorithms that may be used to compute harmonic transforms rapidly.  Numerical experiments have shown our algorithms to be
approximately twice as fast as optimised algorithms implementing the Gauss-Legendre sampling theorem but approximately 25\% slower than the semi-naive algorithm, the most universally applicable algorithm implementing the equiangular Driscoll \& Healy sampling theorem.  However, the semi-naive algorithm applies for scalar functions only, while our sampling theorem also applies directly for spin functions on the sphere (whereas the computation time for a spin \spin\ transform using the semi-naive algorithm would scale by $1+\spin$).  Numerical experiments have also shown our algorithms to be numerically stable to band-limits of $\elmax=4096$.  Conversely, the algorithms that implement the Gauss-Legendre and Driscoll \& Healy sampling theorems on the sphere are restricted in their use of Wigner recursions and, due to the enforced use of the pointwise three-term Wigner recursion, go unstable between band-limits $\elmax=1024$ and $\elmax=2048$.

Our novel sampling theorem and fast algorithms will be of practical benefit both at very high and low band-limits.  For the analysis of \cmb\ data at very high band-limits (\mbox{$\elmax \sim 4096$}), our sampling theorem yields exact spherical harmonic transforms with algorithms that are stable and very accurate.  For the reconstruction of diffusion MRI images at low band-limits (\mbox{$\elmax \sim 10$}), the reduction in the number of samples required by our sampling theorem to represent a band-limited signal may be exploited to reduce the cost of acquisition significantly.  For CS applications in both of these fields and beyond, the reduction in dimensionality and sparsity afforded by our sampling theorem will enhance the performance of CS reconstruction on the sphere.

\bibliographystyle{IEEEtran}
\bibliography{bib}
%\begin{thebibliography}{1}

%H.~Kopka and P.~W. Daly, \emph{A Guide to {\LaTeX}}, 3rd~ed.\hskip 1em plus

%\end{thebibliography}

%\begin{biography}[{\includegraphics[width=1in,height=1.25in,clip,keepaspectratio]{mshell}}]{Michael Shell}
% where an .eps filename suffix will be assumed under latex, and a .pdf suffix
% will be assumed for pdflatex; or if you just want to reserve a space for
% a photo:

\begin{biography}[{\includegraphics[width=1in,height=1.25in,clip,keepaspectratio]%
{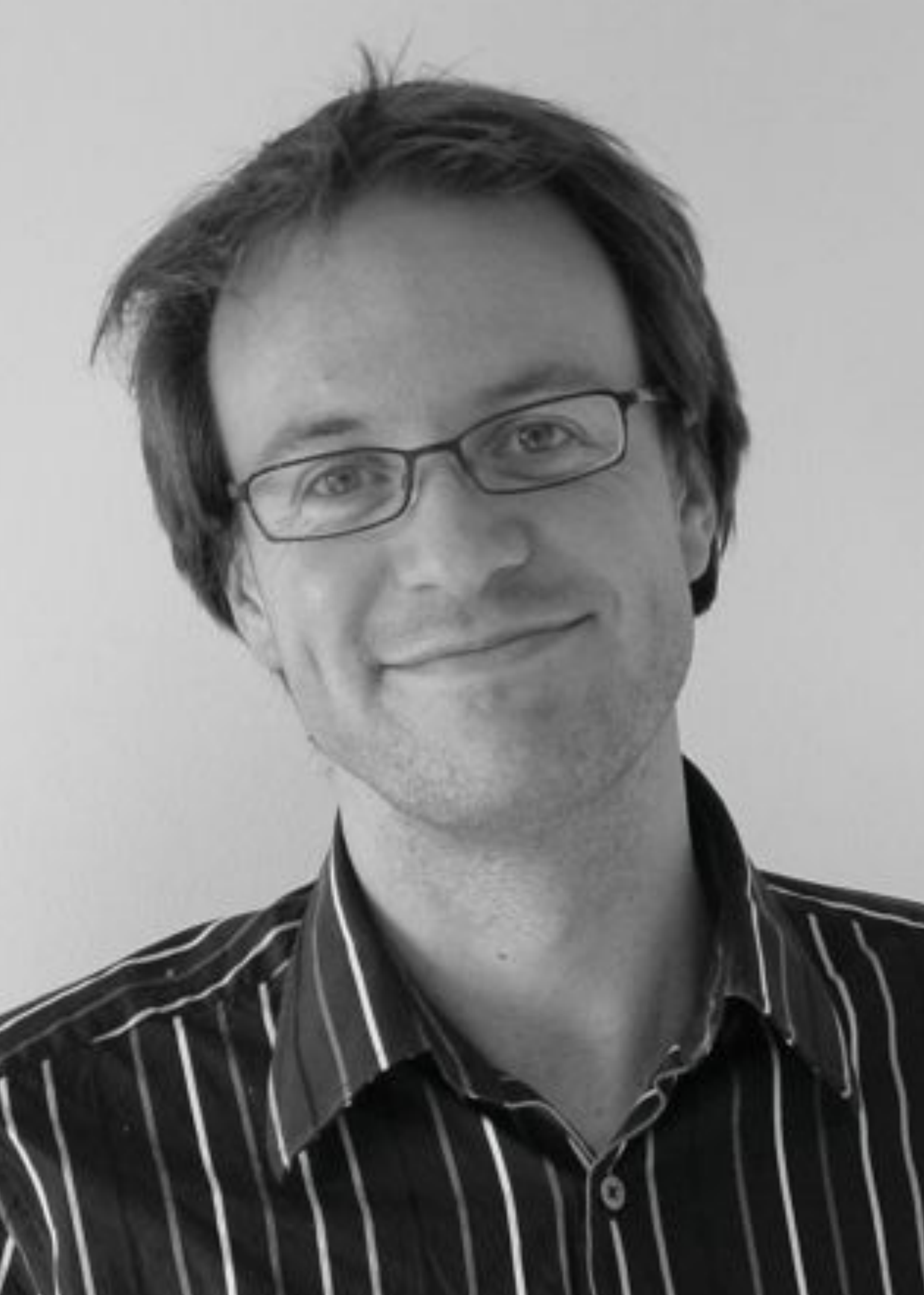}}]{Jason McEwen}
%\begin{biographynophoto}{Jason McEwen}
received a B.E.\ (Hons) degree in Electrical and Computer Engineering
from the University of Canterbury, New Zealand, in 2002 and a Ph.D.\ degree in Astrophysics from the University of Cambridge in 2006. 

He held a Research Fellowship at Clare College, Cambridge, from
2007 to 2008, worked as a Quantitative Analysis in an Investment Bank
from 2008 to 2010, and began a position as a Postdoctoral Researcher at Ecole
Polytechnique F{\'e}d{\'e}rale de Lausanne (EPFL), Switzerland, in
2010.  His research interests include wavelets on the sphere,
compressed sensing, and applications of these theories in cosmology
and radio interferometry.
%\end{biographynophoto}
\end{biography}

\begin{biography}[{\includegraphics[width=1in,height=1.25in,clip,keepaspectratio]{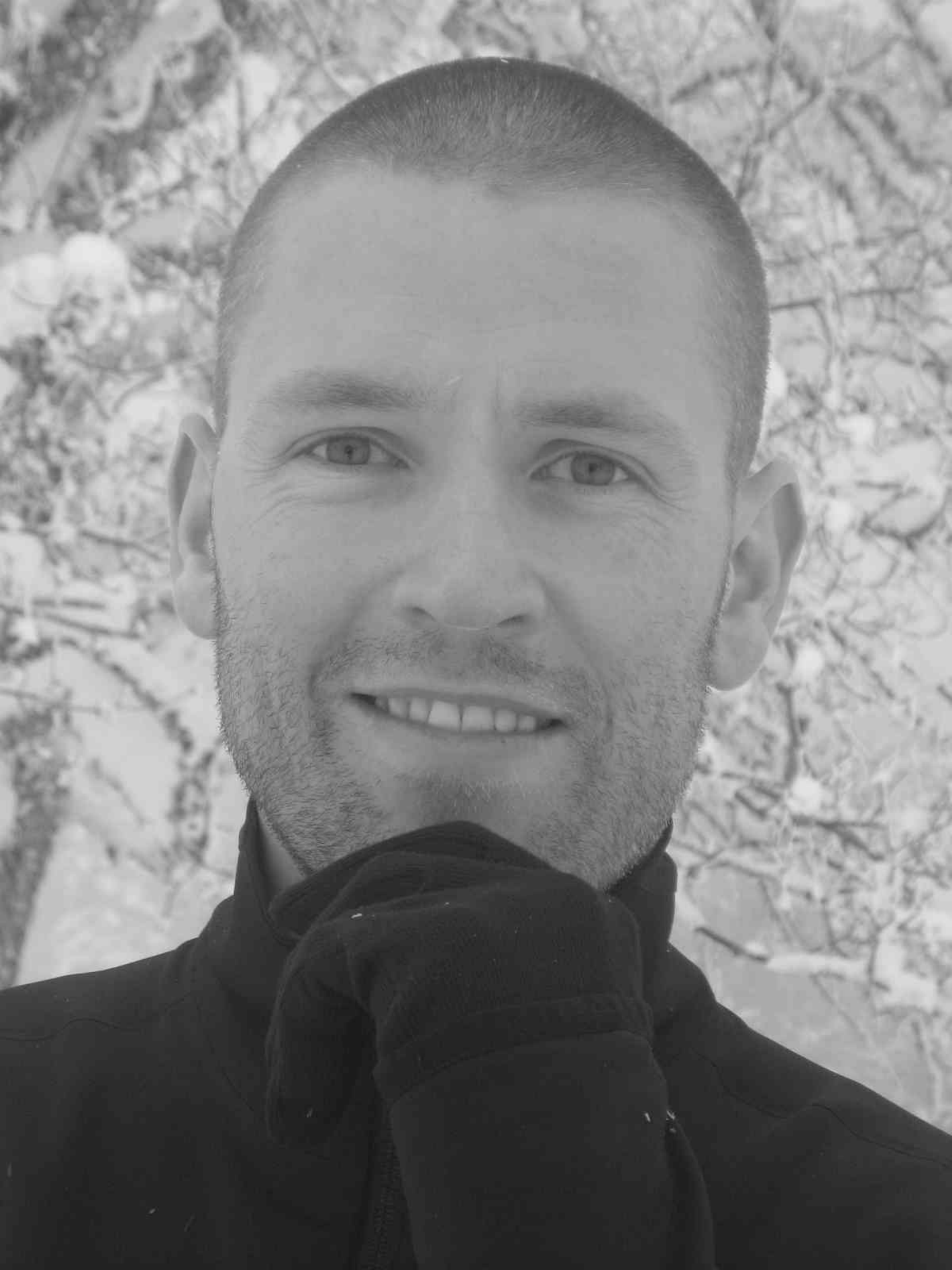}}]{Yves Wiaux}
%\begin{biographynophoto}{Yves Wiaux}
received the M.S. degree in physics and the Ph.D. degree in theoretical physics from the Universit\'e catholique de Louvain (UCL), Louvain-la-Neuve, Belgium, in 1999 and 2002, respectively.

He was a Postdoctoral Researcher at the Signal Processing Laboratories of the Ecole Polytechnique F\'ed\'erale de Lausanne (EPFL), Switzerland, from 2003 to 2008. He was also a Postdoctoral Researcher of the Belgian National Science Foundation (F.R.S.-FNRS) at the Physics Department of UCL from 2005 to 2009. He is now a Ma\^itre Assistant of the University of Geneva (UniGE), Switzerland, with joint affiliation between the Institute of Electrical Engineering and the Institute of Bioengineering of EPFL, and the Department of Radiology and Medical Informatics of UniGE. His research lies at the intersection between complex data processing (including development on wavelets and compressed sensing) and applications in astrophysics (notably in cosmology and radio astronomy) and in biomedical sciences (notably in MRI and fMRI).
%\end{biographynophoto}
\end{biography}

% \vspace*{-10mm}


%\vfill


\end{document}